\documentclass[12pt]{reportj}
\usepackage{deluxetablej}
\usepackage{hyperref} 
\makeatletter
\def\@to{to}
\makeatother
\usepackage{times}
\usepackage{graphicx}
\usepackage{tocloft}
\usepackage{fancyheadings}
\emergencystretch=\maxdimen
\usepackage{datetime}
\newdateformat{ddmonthyyyy}{\THEDAY\ \monthname[\THEMONTH]\ \THEYEAR}

\renewcommand\thesection{\arabic{section}}
\renewcommand\thesubsection{\thesection.\arabic{subsection}}

\def\ssection#1{\setcounter{subsection}{0} \refstepcounter{section} \section*{\hbox to \hsize{\large\bf \arabic{section}. #1\hfill }}\label{sec} \addcontentsline{toc}{section}{\arabic{section}. #1}}
\def\ssubsection#1{\setcounter{subsubsection}{0} \refstepcounter{subsection}\subsection*{\hbox to \hsize{\normalsize\bfseries\itshape \arabic{section}.\arabic{subsection} #1\hfill}}\label{subsec} \addcontentsline{toc}{subsection}{\arabic{section}.\arabic{subsection} #1}}
\def\ssubsubsection#1{\refstepcounter{subsubsection}\subsection*{\hbox to \hsize{\normalsize\it \arabic{section}.\arabic{subsection}.\arabic{subsubsection} #1\hfill}}\label{subsubsec} \addcontentsline{toc}{subsubsection}{\arabic{section}.\arabic{subsection}.\arabic{subsubsection} #1}}

\def\ssectionstar#1{\section*{\hbox to \hsize{\large\bf #1\hfill}} \addcontentsline{toc}{section}{#1}}
\def\ssubsectionstar#1{\subsection*{\hbox to \hsize{\normalsize\bfseries\itshape #1\hfill}} \addcontentsline{toc}{subsection}{#1}}
\def\ssubsubsectionstar#1{\subsection*{\hbox to \hsize{\normalsize\it  #1\hfill}} \addcontentsline{toc}{subsection}{#1}}

\renewcommand{\cftaftertoctitle}{%
\mbox{}\hfill{\normalfont Page}}
\begin{document}

~\\

\vspace{-2.4cm}
\noindent\includegraphics*[width=0.295\linewidth]{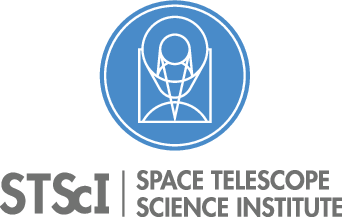}

\vspace{-0.4cm}

\begin{flushright}
    {\bf Instrument Science Report STIS 2026-05(v1)}
    
    \vspace{0.7cm}
    
    {\bf\Huge Spatially Scanned STIS Spectra of the Exoplanet Host Star 55 Cnc}
    
    \rule{0.25\linewidth}{0.5pt}
    
    \vspace{0.4cm}
    
    D. E. Welty$^1$, J. D. Lothringer$^1$, D. K. Sing$^2$, A. M. Jones$^1$, A. Riley$^{1,3}$, and C. R. Proffitt$^1$
    \linebreak
    \newline
    \footnotesize{$^1$ Space Telescope Science Institute, Baltimore, MD\\
                  $^2$ Johns Hopkins University, Dept. of Physics and Astronomy, Baltimore, MD\\
                  $^3$ BAE Systems, Inc., University of Colorado, Boulder, CO\\}
    
    \vspace{0.3cm}
    
    \date{21 August 2026} 
\end{flushright}

\vspace{-0.1cm}

\noindent\rule{\linewidth}{1.0pt}
\noindent{\bf A{\footnotesize BSTRACT}}

{\it \noindent 
We discuss the analysis of two sets of optical/near-IR spectra of the exoplanet host star 55 Cnc, obtained with the Space Telescope Imaging Spectrograph (STIS) and grating G750L in spatial scanning mode, in order to assess the performance of that relatively new observing mode for studies of transiting exoplanets.
Standard pipeline reductions of the CCD spectral images were augmented by custom procedures for removing both cosmic rays and the strong fringing seen at wavelengths longer than about 7000 \AA.
Both total (``white-light'') fluxes and the fluxes for some narrower wavelength intervals were extracted from the processed spectral images.
Apart from slight ($\sim$400 ppm) orbit-to-orbit offsets between the relative fluxes in each set, the patterns exhibited by the flux values within each orbit are very similar.
The systematic differences in the fluxes are somewhat smaller than those seen in archival STIS spectra of 55 Cnc obtained in so-called ``stare mode'', in which the CCD is deliberately saturated at a fixed pointing.
A parameterized detrending method similar to those commonly used to remove instrumental effects from time series observations of exoplanet host stars was then applied to the extracted fluxes.
For the total fluxes, the scatter about the detrending models is of order 30-40 ppm -- comparable to the best precision previously obtained for time series photometry with HST.
The scatter is somewhat larger for the narrower wavelength bins -- particularly at the longer wavelengths where the CCD is less sensitive; the defringing does reduce the scatter by 15-20\% at the longer wavelengths, however.
The depth of the transit of the super-Earth 55 Cnc e ($\sim$ 450 ppm for the total flux) is consistent with previously obtained values.
Both the scan-mode and the stare-mode observations of 55 Cnc e appear to indicate an unexpected (and variable?) increase in the transit radius Rp/Rs between 0.55 and 1.0 $\mu$m (by $>$40\% for the scan-mode data).
While these data are somewhat limited, they do suggest that spatial scanning with the STIS CCD can provide high-quality optical/near-IR spectra of the brighter exoplanet hosts.}

\vspace{-0.1cm}
\noindent\rule{\linewidth}{1.0pt}

\renewcommand{\cftaftertoctitle}{\thispagestyle{fancy}}
\tableofcontents




\lhead{}
\rhead{}
\cfoot{\rm {\hspace{-1.9cm} Instrument Science Report STIS 2026-05(v1) Page \thepage}}

\vspace{-0.3cm}
\ssection{Introduction}\label{sec:Introduction}

For the past ten years, spatial scanning with the Space Telescope Imaging Spectrograph (STIS) onboard the Hubble Space Telescope (HST) has been an available (but unsupported) mode for obtaining high signal-to-noise ratio (S/N) spectra of relatively bright stars with the STIS CCD.
The scanned spectra are obtained by trailing a target along one of the long STIS apertures, perpendicular to the direction of the dispersion.
[Spatial scanning has also been used for observations with WFC3, both to achieve increased observing efficiency and to avoid saturation of bright targets (see, e.g., \href{https://www.stsci.edu/files/live/sites/www/files/home/hst/instrumentation/wfc3/documentation/instrument-science-reports/_documents/2012_08.pdf}{McCullough \& MacKenty 2012}; \href{https://ui.adsabs.harvard.edu/abs/2013ApJ...774...95D/abstract}{Deming et al. 2013}; \href{https://www.stsci.edu/files/live/sites/www/files/home/hst/instrumentation/wfc3/documentation/instrument-science-reports/_documents/2017_21.pdf}{Shanahan et al. 2017}).]
While high-S/N STIS spectra may also be obtained by deliberately saturating the CCD detector at a fixed pointing in so-called ``stare mode'' (e.g., \href{https://ui.adsabs.harvard.edu/abs/1999PASP..111.1009G/abstract}{Gilliland et al. 1999}; \href{https://ui.adsabs.harvard.edu/abs/2004AJ....127.3508B/abstract}{Bohlin \& Gilliland 2004}), spatial scanning offers several distinct advantages.
First, spreading the light over many rows on the CCD detector -- which retains the information about the location of each photon that is lost when charge bleeds along the columns in a saturated image -- should allow better averaging over flat field variations.
Second, the nearly uniform illumination of each CCD column in the scanned spectral image should enable more robust identification and removal of bad pixels and cosmic rays.
Finally, the use of a long slit for both trailed stellar and contemporaneous flat field exposures should yield a better match in the illumination pattern on the detector -- thus enabling more complete removal of the strong fringing seen beyond about 7000 \AA\ in STIS CCD spectra.


\begin{figure}[b!]
\centering
\includegraphics[scale=0.18]{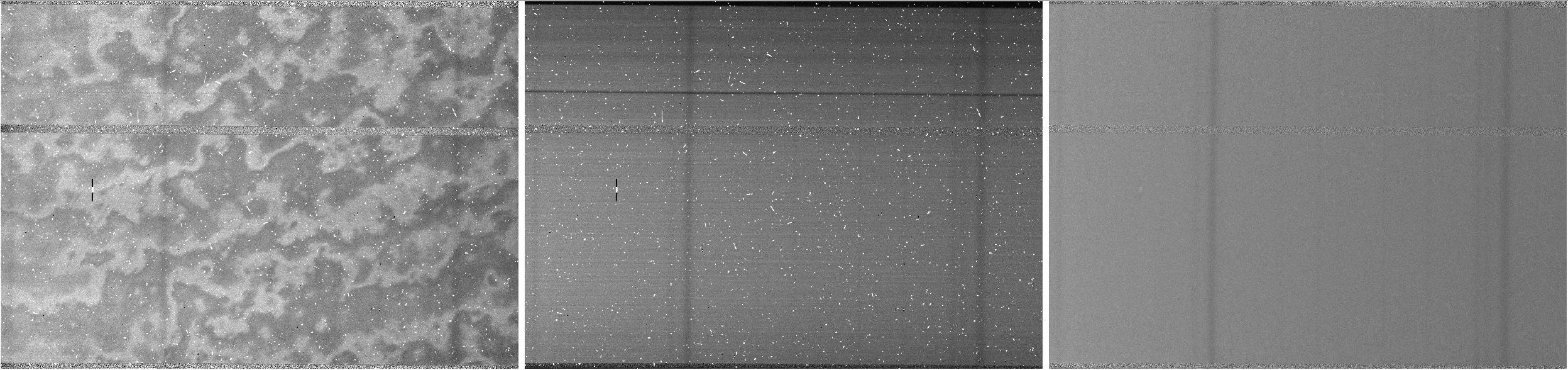}
\caption{{\small Trailed G750M/9336 exposure of a bright star from HST program 14705, illustrating the technique of spatial scanning with the STIS CCD.
In these images, wavelength increases to the right, and the star was scanned in the vertical direction along the length of the narrow 52x0.1 aperture.
At the left is the raw image.
In the center, the fringing pattern has been effectively removed by dividing by a contemporaneous flat-field image, allowing reliable detection of the weak absorption features (vertical dark bands).
Instability in the trail rate and the effects of spacecraft jitter can produce noticeable flux variations as a function of position along the trail direction, however.
At the right, those variations have been removed (along with most cosmic rays and hot pixels).
The corresponding 1D extracted spectrum is shown in \href{https://ui.adsabs.harvard.edu/abs/2017ApJ...843L...2C/abstract}{Cordiner et al. (2017)}.}}
\label{fig:cord}
\end{figure}


Spatial scanning with the STIS CCD can be particularly advantageous for science programs requiring reliable detection and measurement of weak stellar or interstellar absorption in the red and near-IR, where ground-based observations can be severely affected by strong telluric absorption and where the fringing in pointed CCD observations can be difficult to remove.
In the first use of this observing mode, \href{https://ui.adsabs.harvard.edu/abs/2017ApJ...843L...2C/abstract}{Cordiner et al. (2017)} obtained scanned spectra of the heavily reddened B0 I star BD+63$^{\rm o}$1964 with the STIS G750M/9336 setting, in an attempt to detect weak diffuse interstellar bands that have been attributed to C$_{60}^+$ (program 14705).
Use of the 52x0.1 aperture for both the trailed stellar spectra and the tungsten lamp flat field exposures enabled effective removal of the fringing pattern (Figure~\ref{fig:cord}), and the S/N $\sim$ 600-800 achieved in the final summed spectrum is consistent with expectations from the total accumulated counts.
Based on that successful use of spatial scanning, two follow-on programs (15429, 15478) obtained similar scanned spectra of additional targets (\href{https://ui.adsabs.harvard.edu/abs/2019ApJ...875L..28C/abstract}{Cordiner et al. 2019}).

Spatial scanning may also be useful for programs requiring very accurate time series measurements of stellar fluxes (e.g., studies of transiting exoplanets and their atmospheres).
There is a rapidly growing literature on studies of transiting exoplanets undertaken with HST -- generally using either pointed/saturated spectra obtained with STIS or spatially scanned spectra obtained with the WFC3 grisms (e.g., \href{https://ui.adsabs.harvard.edu/abs/2001ApJ...552..699B/abstract}{Brown et al. 2001}; \href{https://ui.adsabs.harvard.edu/abs/2011A&A...527...73S/abstract}{Sing et al. 2011}, \href{https://ui.adsabs.harvard.edu/abs/2019AJ....158...91S/abstract}{2019}; \href{https://ui.adsabs.harvard.edu/abs/2012MNRAS.422.2477H/abstract}{Huitson et al. 2012}; \href{https://ui.adsabs.harvard.edu/abs/2013ApJ...772L..16E/abstract}{Evans et al. 2013}; \href{https://ui.adsabs.harvard.edu/abs/2014MNRAS.437...46N/abstract}{Nikolov et al. 2014}; \href{https://ui.adsabs.harvard.edu/abs/2015MNRAS.450.2043D/abstract}{Demory et al. 2015}; \href{https://ui.adsabs.harvard.edu/abs/2016ApJ...819...10W/abstract}{Wakeford et al. 2016}; \href{https://ui.adsabs.harvard.edu/abs/2016ApJ...832..202T/abstract}{Tsiaras et al. 2016}).
Those studies typically find systematic trends in the observed stellar fluxes, both between orbits in a given visit and within each individual orbit, that are thought to be due to instrumental/observational effects.
Those trends -- which can vary from object to object and from visit to visit and which can range in amplitude from several hundred to several thousand ppm -- must be understood and removed in order to obtain both accurate measurements of the exoplanetary transits and subsequent constraints on the composition and properties of exoplanetary atmospheres.
Linear regression fits to the observed time series fluxes -- parameterized by time, HST orbital phase, precise location of the spectral image on the detector (and perhaps other factors) and generally including models for stellar limb darkening and the transit profile -- thus are commonly employed to try to discern the underlying physical variations in the fluxes.
While those empirical ``detrending'' fits have enabled various studies to achieve rms scatter (versus the fits) as low as several tens of ppm in the total ``white-light'' fluxes (within a given visit to a particular target), the observed differences in the trends from visit to visit have hindered comparisons of the data over longer time periods.

This report presents the analysis of images and spectra of the exoplanet host star 55 Cnc obtained for STIS special calibration programs 15383 (C. Proffitt, PI) and 16442 (D. Welty, PI).
The two programs were designed to test the ability of spatial scanning with the STIS CCD to perform very accurate measurement and monitoring of stellar fluxes -- both broad band and in narrower spectral intervals -- as needed (for example) for characterizing transiting exoplanets and their atmospheres.
The results derived from these scan-mode observations are compared with those obtained from earlier stare-mode observations of the same target (program 13665; B. Benneke, PI), which used the same instrumental configuration.
Section~\ref{sec-obs} describes the data obtained for the three observing programs.
Section~\ref{sec-proc} discusses the procedures used to identify and remove the effects of cosmic rays, to remove the fringing seen at longer wavelengths, and to extract 1D spectra from the 2D spectral images.
Section~\ref{sec-detr} gives a brief overview of the detrending procedure used to remove the effects of various instrumental/observational systematics.
Section~\ref{sec-res} discusses the results regarding scan geometry, photometric stability, and the transit characteristics obtained from the data.
Section~\ref{sec-sum} summarizes those results and makes some recommendations for when and how to use spatial scans with STIS.
Two appendices provide information on various exposure parameters (and the adopted detrending coefficients), and on the jitter behavior during the scans of 55 Cnc.
 
%
\begin{deluxetable}{ccccccccl}
\tablecolumns{9}
\tabletypesize{\scriptsize}
\tablecaption{Special STIS calibration program 15383 \label{tab:obs}}
\tablewidth{0pt}

\tablehead{
\multicolumn{1}{c}{Visit}&
\multicolumn{1}{c}{Orbit}&
\multicolumn{1}{c}{Dataset}&
\multicolumn{1}{c}{Target}&
\multicolumn{1}{c}{Aperture}&
\multicolumn{1}{c}{t$_{\rm exp}$}&
\multicolumn{1}{c}{Scan Rate}&
\multicolumn{1}{c}{Length}&
\multicolumn{1}{c}{Description}\\
\multicolumn{1}{c}{ }&
\multicolumn{1}{c}{ }&
\multicolumn{1}{c}{ }&
\multicolumn{1}{c}{ }&
\multicolumn{1}{c}{ }&
\multicolumn{1}{c}{(s)}&
\multicolumn{1}{c}{(arcsec/s)}&
\multicolumn{1}{c}{(arcsec)}&
\multicolumn{1}{c}{ }}

\startdata
 1 & 1 & 01bjq       & W lamp      & 52x0.1 & 0.5   & ...      & ... & lamp image \\
   &   & 01bkq-01blq & GRW+70 5824 & 52x2   & 100   & 0.5      &  50 & trailed stellar images \\
   &   & 01bmq       & W lamp      & 52x0.1 & 0.5   & ...      & ... & lamp image \\
   &   & 01bnq-01bpq & GRW+70 5824 & 52x2   & 100   & 0.5      &  50 & trailed stellar images \\
   &   & 01bqq       & W lamp      & 52x0.1 & 0.5   & ...      & ... & lamp image \\
   &   & 01brq-01bsq & GRW+70 5824 & 52x2   & 100   & 0.5      &  50 & trailed stellar images \\
   &   & 01bvq       & W lamp      & 52x0.1 & 0.5   & ...      & ... & lamp image \\
   &   & 01bwq-01bxq & GRW+70 5824 & 52x2   & 100   & 0.5      &  50 & trailed stellar images \\
   &   & 01byq       & W lamp      & 52x0.1 & 0.5   & ...      & ... & lamp image \\
   &   & \\
 2 & 1 & 11010       & W lamp      & 52x0.1 & 2x25  & ...      & ... & default fringe flats \\
   &   & 11taq-11tbq & 55 Cnc      & 52x0.1 & 368   & 0.130435 &  48 & long scans (narrow slit)  \\
   &   & 11tcq-11tdq & 55 Cnc      & 52x2   & 368   & 0.130435 &  48 & long scans (wide slit)    \\
   &   & 11020       & W lamp      & 52x0.1 & 2x25  & ...      & ... & default fringe flats \\
   &   & 11040       & W lamp      & 52x0.1 & 5x210 & ...      & ... & deep fringe flats    \\
   & 2 & 11trq-11u1q & 55 Cnc      & 52x2   & 218   & 0.055    &  12 & short scans (sub-array) \\
   &   & 11050       & W lamp      & 52x0.1 & 2x25  & ...      & ... & default fringe flats \\
   &   & 11060       & W lamp      & 52x0.1 & 6x210 & ...      & ... & deep fringe flats    \\
   & 3 & 11ueq-11uoq & 55 Cnc      & 52x2   & 218   & 0.055    &  12 & short scans (sub-array) \\
   &   & 11070       & W lamp      & 52x0.1 & 2x25  & ...      & ... & default fringe flats \\
   &   & 11080       & W lamp      & 52x0.1 & 6x210 & ...      & ... & deep fringe flats    \\
\enddata
\tablecomments{All dataset names begin with 'odqf'; all exposures in visit 2 used setting G750L/7751.
Default wavecals (52x0.1, 6.2 s) were obtained before and after the stellar scans in visit 2.
The deep fringe flats in visit 2 were obtained during occultation.}
\end{deluxetable}

\begin{deluxetable}{ccccccccl}
\tablecolumns{9}
\tabletypesize{\scriptsize}
\tablecaption{Special STIS calibration program 16442 \label{tab:obs2}}
\tablewidth{0pt}

\tablehead{
\multicolumn{1}{c}{Visit}&
\multicolumn{1}{c}{Orbit}&
\multicolumn{1}{c}{Dataset}&
\multicolumn{1}{c}{Target}&
\multicolumn{1}{c}{Aperture}&
\multicolumn{1}{c}{t$_{\rm exp}$}&
\multicolumn{1}{c}{Scan Rate}&
\multicolumn{1}{c}{Length}&
\multicolumn{1}{c}{Description}\\
\multicolumn{1}{c}{ }&
\multicolumn{1}{c}{ }&
\multicolumn{1}{c}{ }&
\multicolumn{1}{c}{ }&
\multicolumn{1}{c}{ }&
\multicolumn{1}{c}{(s)}&
\multicolumn{1}{c}{(arcsec/s)}&
\multicolumn{1}{c}{(arcsec)}&
\multicolumn{1}{c}{ }}

\startdata
 1 & 1 & 01kuq-01kzq & 55 Cnc      & 52x2   & 192   & 0.0625 &  12 & short scans (sub-array) \\
   &   & 01010       & W lamp      & 52x0.1 & 2x25  & ...    & ... & default fringe flats \\
   &   & 01020       & W lamp      & 52x0.1 & 5x210 & ...    & ... & deep fringe flats    \\
   & 2 & 01lcq-01llq & 55 Cnc      & 52x2   & 192   & 0.0625 &  12 & short scans (sub-array) \\
   &   & 01030       & W lamp      & 52x0.1 & 2x25  & ...    & ... & default fringe flats \\
   &   & 01040       & W lamp      & 52x0.1 & 6x210 & ...    & ... & deep fringe flats    \\
   & 3 & 01lxq-01m6q & 55 Cnc      & 52x2   & 192   & 0.0625 &  12 & short scans (sub-array) \\
   &   & 01050       & W lamp      & 52x0.1 & 2x25  & ...    & ... & default fringe flats \\
   &   & 01060       & W lamp      & 52x0.1 & 6x210 & ...    & ... & deep fringe flats    \\
   & 4 & 01mjq-01msq & 55 Cnc      & 52x2   & 192   & 0.0625 &  12 & short scans (sub-array) \\
   &   & 01070       & W lamp      & 52x0.1 & 2x25  & ...    & ... & default fringe flats \\
   &   & 01080       & W lamp      & 52x0.1 & 6x210 & ...    & ... & deep fringe flats    \\
   & 5 & 01n4q-01ndq & 55 Cnc      & 52x2   & 192   & 0.0625 &  12 & short scans (sub-array) \\
   &   & 01090       & W lamp      & 52x0.1 & 2x25  & ...    & ... & default fringe flats \\
   &   & 010a0       & W lamp      & 52x0.1 & 6x210 & ...    & ... & deep fringe flats    \\
\enddata
\tablecomments{All dataset names begin with 'oeja'; all exposures used setting G750L/7751.
Default wavecals (52x0.1, 6.2 s) were obtained before and after the stellar scans.
The deep fringe flats were obtained during occultation.}
\end{deluxetable}

\ssection{Observations}\label{sec-obs}

The well studied bright, nearby K0 IV-V star 55 Cnc ($V$ = 5.95; d = 12.6 pc) exhibits both a stellar activity cycle (with period $\sim$10.5 yr) and rotational variations in flux (with period $\sim$38.8 d and amplitude $\sim$3000 ppm; \href{https://ui.adsabs.harvard.edu/abs/2018A&A...619....1B/abstract}{Bourrier et al. 2018}).
Of the five known exoplanets hosted by 55 Cnc, the close-in super-Earth 55 Cnc e (M $\sim$ 8 M$_{\rm{Earth}}$; semi-major axis $\sim$ 0.0155 AU; period $\sim$ 17.7 hr) exhibits primary transits lasting for about 1.6 hr, with transit depth $\sim$ 400 ppm (\href{https://ui.adsabs.harvard.edu/abs/2008ApJ...675..790F/abstract}{Fischer et al. 2008}; \href{https://ui.adsabs.harvard.edu/abs/2010ApJ...722..937D/abstract}{Dawson \& Fabrycky 2010}; \href{https://ui.adsabs.harvard.edu/abs/2011ApJ...737L..18W/abstract}{Winn et al. 2011}; \href{https://ui.adsabs.harvard.edu/abs/2018A&A...619....1B/abstract}{Bourrier et al. 2018}).
Some observed out of transit variations in the optical/NIR flux, apparently with the same periodicity as the orbit of 55 Cnc e, may be due (at least in part) to thermal emission from the planet -- though the amplitude ($\sim$30 to $\sim$200 ppm), phase, and shape of the variations appear to depend on both the epoch and passband of the observations (\href{https://ui.adsabs.harvard.edu/abs/2019A&A...631..129S/abstract}{Sulis et al. 2019}).
55 Cnc e is the focus of several JWST programs aimed at characterizing the surface and atmospheric properties of that hot, rocky exoplanet.

As detailed in Tables~\ref{tab:obs} and \ref{tab:obs2}, program 15383 included two visits, which were executed on 2017 Nov 06 and 2017 Dec 28, and program 16442 included a single visit, executed on 2021 Apr 25.
The first visit of program 15383 was designed to test the scan geometry, by comparing long ``round-trip'' imaging scans of the white dwarf GRW+70 5824 (taken through the 52x2 aperture at a nominal scan angle of 90 degrees) with images of the onboard tungsten lamp (taken through the 52x0.1 aperture).
The second visit of program 15383 was designed to test the reproducibility of the fluxes -- both within a given orbit and from orbit to orbit -- using a series of 20 identical short scans of 55~Cnc, taken through the 52x2 aperture.
That second visit was scheduled to avoid the transit of 55 Cnc e, in order to minimize variations in the stellar flux during the visit.
Program 16442 obtained a longer time series of scanned spectra -- 42 scans over five orbits, covering a transit of 55 Cnc e -- to see how well the transit depth could be characterized.

\begin{figure}[b!]
\centering
\includegraphics[scale=0.8]{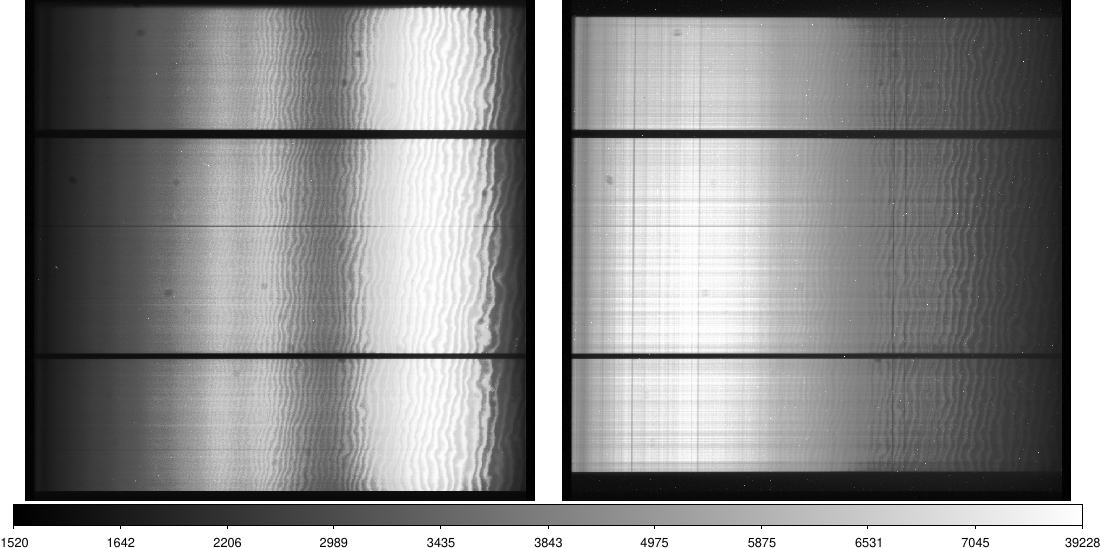}
\caption{{\small ({\it left}) Raw G750L/7751 image of the tungsten lamp, taken through the 52x0.1 aperture.
({\it right}) Raw G750L/7751 spectral image of 55 Cnc from program 15383, trailed along the 52x0.1 aperture.
Wavelength (x axis) increases to the right; note the broad horizontal occulting bars and the fringing at the longer wavelengths.
In the 55 Cnc spectral image, the narrow vertical dark lines are due to stellar absorption; the horizontal bands may reflect spacecraft jitter and/or variations in the scan rate and/or the aperture width.}}
\label{fig:rawfull}
\end{figure}

Previous time series observations with HST covering multiple orbits have shown that data obtained in the first orbit of a visit often exhibit differences in behavior, relative to the rest of the orbits.
Those differences are thought to be related to the telescope ``settling into'' the thermal environment of its new pointing (e.g., \href{https://ui.adsabs.harvard.edu/abs/2001ApJ...552..699B/abstract}{Brown et al. 2001}; \href{https://ui.adsabs.harvard.edu/abs/2012MNRAS.422.2477H/abstract}{Huitson et al. 2012}).
The first orbit of the second visit of program 15383 was therefore devoted to several long (48 arcsec) G750L/7751 scans of 55 Cnc through the 52x0.1 and 52x2 apertures, covering the wavelength range from about 5240 to 10270 \AA\ at a resolution of about 500.
These observations were obtained to assess how well the fringing at wavelengths beyond about 7000 \AA\ can be corrected in scanned spectral images taken through the wider slits employed when accurate fluxes are desired.
(The tungsten lamp flat field images used for defringing are generally taken through the narrow 52x0.1 slit, in order to better match the illumination of the detector by a point source in the wider slits.)
Figure~\ref{fig:rawfull} compares one of the long G750L fringe flat exposures ({\it left}) with one of the long scans of 55 Cnc ({\it right}); both images were taken through the 52x0.1 slit.
Although the spectral shapes are different, both images exhibit the characteristic fringing at wavelengths longer than about 7000 \AA\ (with variations primarily along the dispersion direction for G750L), the shadows of the two horizontal occulting bars, and narrower, nearly horizontal stripes.
The stripes in common to both images may be due to slight variations in the width of the slit, while the additional stripes seen in the 55 Cnc spectral image may reflect a combination of spacecraft jitter and variations in the scan rate.

\begin{figure}[b!]
\centering
\includegraphics[scale=0.34]{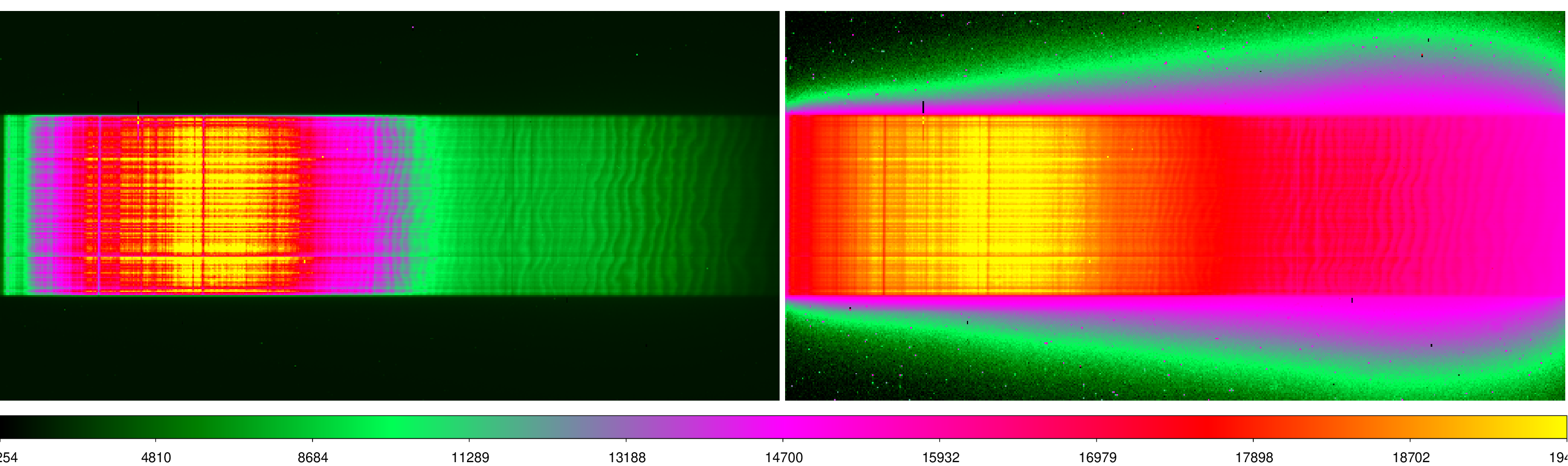}
\caption{{\small A minimally processed G750L/7751 scanned 512x1024 sub-array image of 55 Cnc from program 15383, shown with linear and log stretches to emphasize the structure and features in the main part of the scan ({\it left}) or the weaker features outside the scanned region ({\it right}).
The image has been corrected for bias and dark counts; standard low-order and pixel-to-pixel flat fields have been applied.
Wavelength (x axis) increases to the right; the color bar refers to the left-hand image only.
Note the weak scattered light halo (stronger at longer wavelengths) and the many cosmic rays in the background regions in the right-hand image.}}
\label{fig:fltsub}
\end{figure}

The second and third orbits of the second visit of program 15383 were then used to obtain two sets of 10 identical short scanned G750L/7751 spectra of 55 Cnc (through the 52x2 aperture), to assess the stability of the fluxes over those two orbits.
A 512x1024 sub-array was used for those short scans, in order to increase the efficiency of the observations by reducing the overhead due to CCD readout.
The target was positioned at the same position (6 arcsec below the nominal center of the CCD) at the beginning of each scan, and the adopted 0.055 arcsec s$^{-1}$ scan rate (corresponding to $\sim$ 1.1 pix s$^{-1}$) yielded a total spatial extent of 12 arcsec for each 218-second scanned exposure.
Each exposure exhibits a well exposed, fairly uniform ``plateau'' occupying about 240 CCD rows (near the center of the 512 rows in the sub-array; Figure~\ref{fig:fltsub}), with short ``ramps'' above and below that plateau (reflecting the stellar psf and reduced effective exposure time near the beginning and end of the scan).
The short sub-array scans thus fall in between the shadows of the two occulting bars on the CCD, and they exhibit the same stellar lines, fringing, and (variable) horizontal striping seen in the longer scans that were obtained during the first orbit. 
There is also a weak scattered light halo (especially at longer wavelengths; right-hand panel of Fig.~\ref{fig:fltsub}).
The total counts are of order 1-2 $\times$ 10$^4$ per pix (and thus $\sim$ 2.5-5.0 $\times$ 10$^6$ per column or wavelength bin) over the shorter wavelength half of the plateau region, but decline somewhat at higher wavelengths, due to the reduced sensitivity of the CCD there.
For G750L, the primary variations of the fringing are along the dispersion direction, with amplitudes up to $\pm$10-15\% and with the maxima (or minima) typically separated by 15-20 pix.

All five orbits in the single visit of program 16442 were filled with similar short scans of 55 Cnc with G750L/7751 -- but using a slightly faster scan rate (0.0625 arcsec s$^{-1}$) and shorter exposure time (192 sec) in order to accommodate increased instrumental overheads.
Those scan parameters yielded a very similar overall spatial extent for the scans to that obtained in program 15383, but somewhat lower total counts per pixel in the individual spectral images.
In both programs, the individual scans were placed very consistently on the CCD detector, within several tenths of a pixel in all cases.

\begin{figure}[b!]
\centering
\includegraphics[scale=0.16, clip = true, trim = 0 500 0 500]{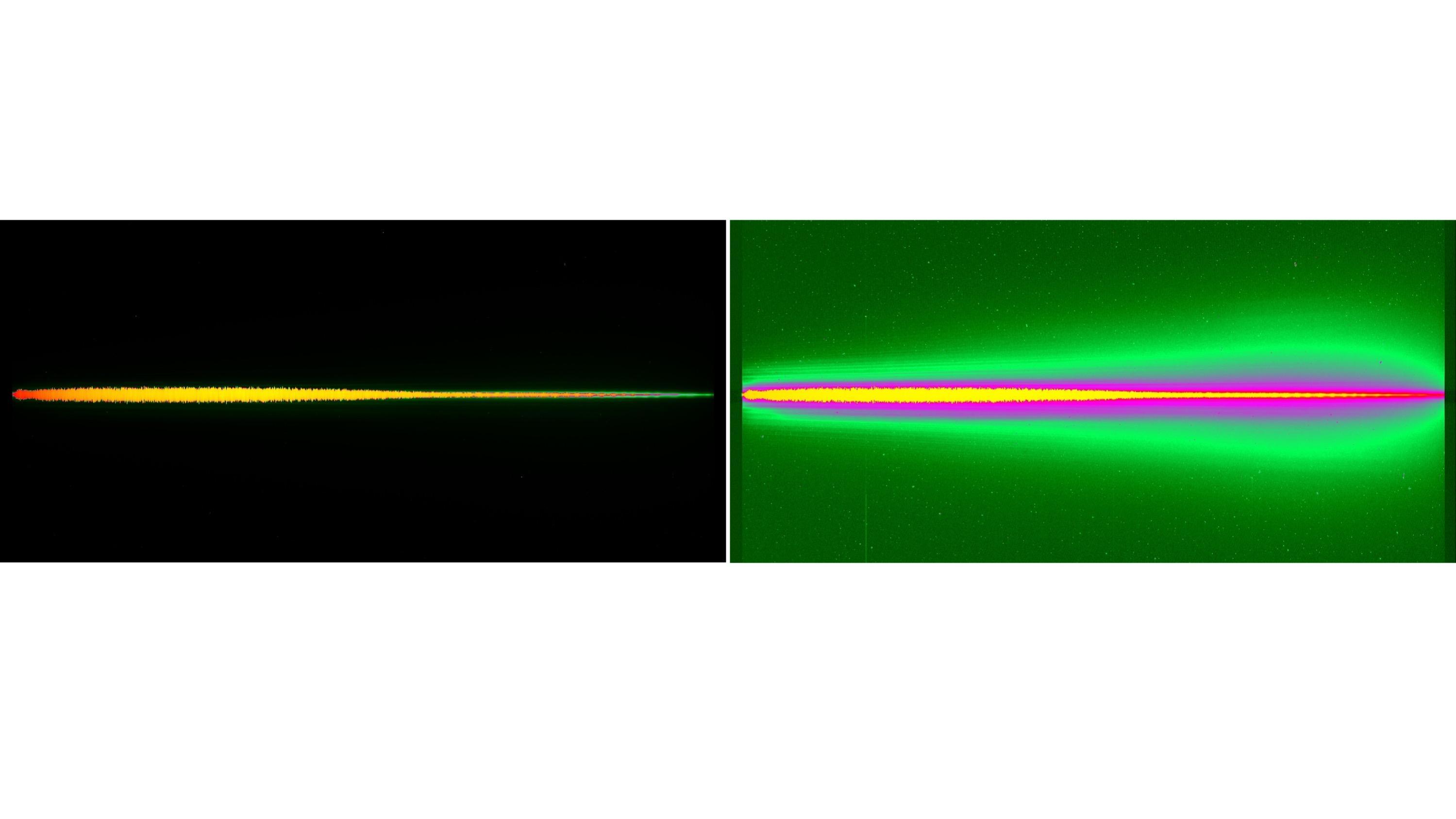}
\caption{{\small A raw G750L/7751 stare-mode sub-array image of 55 Cnc from program 13665, shown with linear and log color stretches to emphasize the structure and features in the main, largely saturated part of the image ({\it left}) or the weaker features outside the main region ({\it right}).
Wavelength (x axis) increases to the right in each image; the maximum y extent of the saturated part of the spectrum (in yellow) is of order 25 pix (much larger than the default 7-pix extraction box).
Note the weak scattered light halo (stronger at longer wavelengths) and the cosmic rays in the background regions in the right-hand image.}}
\label{fig:rawstare}
\end{figure}

In the usual stare mode, 55 Cnc (V=5.95) saturates the STIS CCD in only 1.16 seconds, which is much less time than it takes to read out the detector.
In order to increase observational efficiency, longer exposures are necessary --  either deliberately saturating the detector beyond the full well depth of the pixels (in stare mode) or avoiding saturation by performing spatial scans.
To compare the results of those two approaches, we also analyzed archival data from program 13665 (B. Benneke, PI), which observed three separate transits of 55 Cnc e with STIS G750L in stare mode, on 2014 Oct 30, 2015 May 10, and 2015 May 22 (visits 13, 14, 15 -- which we will refer to as visits 1, 2, 3).
Because the target, instrument, and grating were the same as for our spatially scanned observations, meaningful comparisons can be made between the outcomes of those two very different observing strategies.
The stare-mode observations of 55 Cnc in program 13665 were each 36 seconds long -- i.e., approximately 31 times beyond saturation.
Figure~\ref{fig:rawstare} shows one of the saturated stare-mode images, with different color stretches to show the structure in the 2D image.
While the expectation has been that electrons overflowing the full wells in such overexposures will bleed conservatively along columns -- thus preserving linearity of response, a reanalysis of some saturated STIS CCD exposures has identified serial artifacts appearing beyond saturation, potentially violating that assumption of linearity (\href{https://www.stsci.edu/files/live/sites/www/files/home/hst/instrumentation/stis/documentation/instrument-science-reports/_documents/2015_06.pdf}{Proffitt 2015}).
Our spatial scan dataset therefore provides an important test of the behavior of highly saturated STIS CCD observations by providing a non-saturated control observation of the same target.
Analysis of a similar, but less saturated dataset from program 13665 of the transit of the sub-Neptune HD 97658b was presented in \href{https://ui.adsabs.harvard.edu/abs/2020AJ....159..239G/abstract}{Guo et al. (2020)}.


\ssection{Data processing}\label{sec-proc}

Processing of the scanned spectral images began with standard {\bf calstis} corrections for bias, flat fielding (both low order and pixel-to-pixel), and dark counts.
Given the counts accumulated in the exposures (of order 10$^4$ per pixel in the plateau region), the effects of charge transfer inefficiency (CTI) were expected to be insignificant (e.g., \href{https://www.stsci.edu/files/live/sites/www/files/home/hst/instrumentation/stis/documentation/instrument-science-reports/_documents/2006_03.pdf}{Goudfrooij \& Bohlin 2006}); CTI-induced trails were seen only for strong cosmic rays in the background region below the plateau region, but not within the plateau or in the background region above the plateau.
Following the initial processing, custom procedures were used to identify and correct pixels affected by cosmic rays and/or ``bad'' pixels that were incompletely corrected in the flat fielding.
The effects of fringing were then removed via a recently developed \href{https://stistools.readthedocs.io/en/latest/defringe.html}{Python package} modeled on the previously used {\sc iraf} defringing routines.
Finally, a detrending procedure similar to those commonly employed in analyses of time series data for exoplanet host stars (e.g., \href{https://ui.adsabs.harvard.edu/abs/2011A&A...527...73S/abstract}{Sing et al. 2011}, \href{https://ui.adsabs.harvard.edu/abs/2019AJ....158...91S/abstract}{2019}) was used to fit the 55 Cnc fluxes extracted from the defringed scanned spectral images.


\ssubsection{Cosmic ray removal}\label{sec-cr}

STIS CCD spectra taken at a fixed pointing normally are obtained in two or more identical exposures (with crsplit or repeatobs greater than 1), so that cosmic rays may be identified and corrected by comparing those multiple exposures (e.g., with {\bf ocrreject}; \href{https://www.stsci.edu/files/live/sites/www/files/home/hst/instrumentation/stis/documentation/instrument-science-reports/_documents/1998_22.pdf}{Shaw \& Hodge 1998}).
Because transiting exoplanet time-series observations often consist of many repeated exposures, cosmic rays have typically been identified in the time domain, then removed using customized procedures designed to avoid over-flagging and over-subtraction (e.g., \href{https://ui.adsabs.harvard.edu/abs/2014MNRAS.437...46N/abstract}{Nikolov et al. 2014}).
The scanned STIS exposures, however, were obtained with crsplit=no, and the differences in striping among the individual exposures made it difficult to reliably identify cosmic rays (particularly the weaker ones) by intercomparing the exposures.
It was therefore necessary to rely on spatial information within each individual spectral image to flag and remove the pixels affected by cosmic rays.

Two different procedures for cosmic ray cleaning were explored.
In both cases, the first step is to ``de-stripe'' the images by normalizing each row in the uniformly exposed plateau region -- scaling each one with respect to the median of the counts in columns 100-400 in that region (where fringing is not an issue).
While the striping is not strictly along the rows, but rather along the spectral traces, both the fiducial G750L traces and the observed striping deviate from the rows by only several tenths of a pixel across the CCD.
Normalizing the rows was thus quite sufficient for the purpose of identifying the cosmic rays, particularly for the shorter sub-array scans.
The normalization was not performed for the narrow ``ramp'' regions above and below the plateau or for the background regions.

In the first approach, the median and standard deviation for the plateau region were computed for each column in the de-striped image, and pixels deviating by more than 3$\sigma$ were flagged and replaced in the original (striped) image by the median value (scaled by the de-striping factor for that pixel's row).
While the lines of constant wavelength are not exactly along the columns, they vary by only about 1.5 pix over the plateau region in the short sub-array scans, so that taking the median along the columns was adequate for identifying the cosmic rays.
That identification/correction step was then repeated once for the sub-array scans and twice for any longer full-frame scans that were obtained.
While this procedure was able to identify most of the clear cosmic ray hits, some residual structures from the de-striping were also identified as cosmic rays, resulting in over-flagging.

The second approach to cosmic ray removal employed the well-known package L.A.C{\sc osmic} (\href{https://ui.adsabs.harvard.edu/abs/2001PASP..113.1420V/abstract}{van Dokkum 2001}), which was designed to remove cosmic rays of arbitrary shape from individual images via Laplacian edge detection.
Optimal conservative parameters for L.A.C{\sc osmic} were found through trial and error, with the final contrast (i.e., the contrast threshold between the Laplacian image and the fine-structure image) set to 2, an initial cosmic-ray threshold of 12$\sigma$, and a neighbor threshold of 10$\sigma$ (i.e., if a pixel neighbors another pixel already identified as a cosmic ray, it is treated with a slightly lower threshold).
With that fine-tuning, L.A.C{\sc osmic} was able to identify and remove more of the actual cosmic rays, without flagging the de-striping residuals -- and so was adopted for processing the scanned data.


\ssubsection{Defringing}\label{sec-defr}

\begin{figure}[t!]
\centering
\includegraphics[scale=0.4]{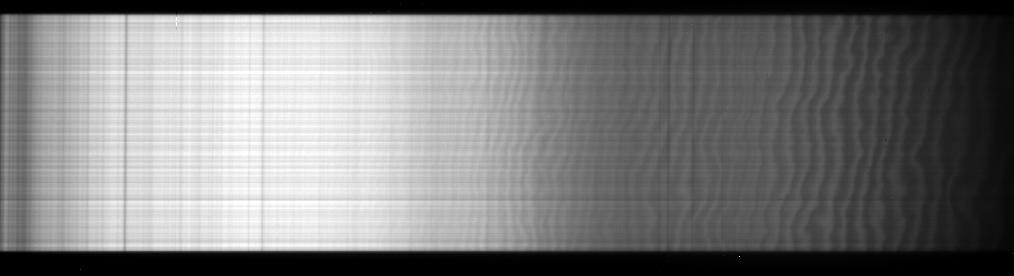}
\includegraphics[scale=0.4]{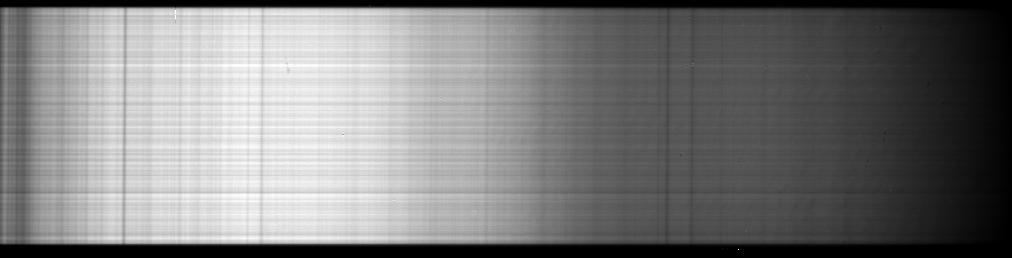}
\caption{{\small One of the 20 short scanned G750L spectral images of 55 Cnc from program 15383.
Wavelength increases from left to right; the scan is performed roughly perpendicular to the dispersion direction.
At the top is the cosmic-ray-corrected image, with the fringing pattern apparent at longer wavelengths; at the bottom is the defringed image.
In both cases, the vertical lines are stellar absorption features, while the horizontal “striping” (which differs from image to image) is likely due to a combination of spacecraft jitter and (perhaps) slight variations in the scan rate.}}
\label{fig:defr}
\end{figure}

Fringing is apparent in both STIS G750M and G750L spectral images at wavelengths greater than about 7000 \AA.
For G750M, the fringing pattern has structure in both the dispersion and spatial directions (Fig.~\ref{fig:cord}).
For G750L, the primary variations of the fringing are along the dispersion direction, with amplitudes up to $\pm$10-15\% and with the maxima (or minima) typically separated by 15-20 pix (Figs.~\ref{fig:rawfull}, \ref{fig:fltsub}, and \ref{fig:defr}).
For spectra taken at a fixed pointing, the current {\bf calstis} pipeline reductions do not correct the fringing, as the defringing process can require some tuning for each individual image -- even when using the recommended contemporaneous fringe flat exposures.
The effects of the uncorrected fringing are usually quite obvious in the extracted spectra.
Those effects are reduced for spectra extracted from spatially scanned spectral images, due to averaging over a wider range in the spatial direction at each wavelength, and they can be essentially eliminated if explicit defringing of the 2D spectral image is performed (e.g., fig.~1 in \href{https://ui.adsabs.harvard.edu/abs/2017ApJ...843L...2C/abstract}{Cordiner et al. 2017}).
While defringing may not be critical for time series analyses of very broad-band fluxes, it becomes increasingly important for considerations of the fluxes in narrower spectral bands (e.g., \href{https://ui.adsabs.harvard.edu/abs/2014MNRAS.437...46N/abstract}{Nikolov et al. 2014}).

\begin{figure}[t!]
\centering
\includegraphics[scale=0.15, clip = true, trim = 0 250 0 250]{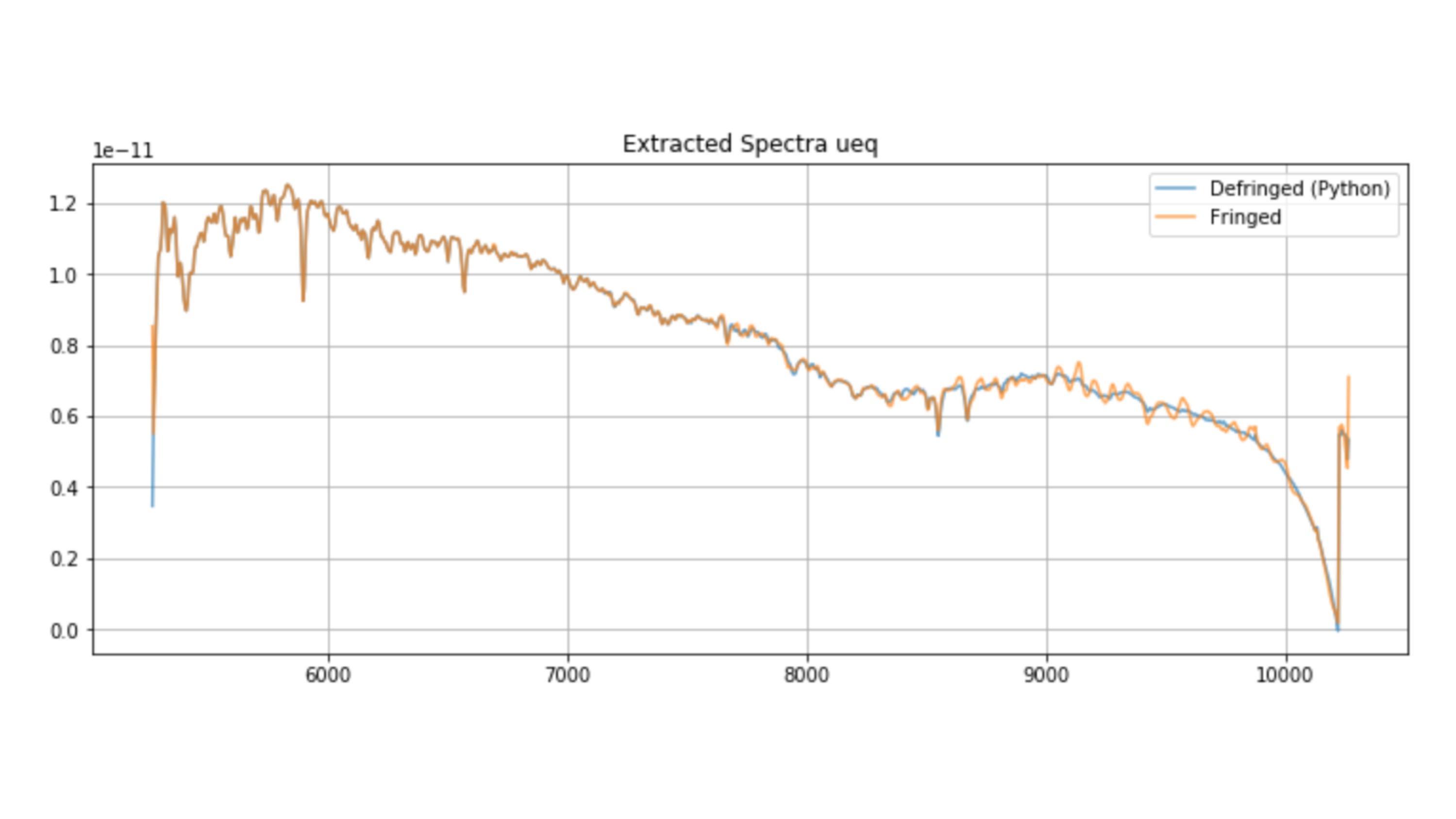}
\caption{{\small An extracted spectrum of 55 Cnc from program 15383, before and after defringing.
While the fringing at wavelengths above 7000 \AA\ is reduced by summing over the full $\sim$260-pix extent of the scan (yellow curve), it is not completely eliminated (e.g., above 8500 \AA).
The flux in the defringed spectrum (blue curve) agrees very well with that in the uncorrected sum.}}
\label{fig:extr}
\end{figure}

Correction of the fringing in our spatially scanned STIS G750L spectral images was performed with a recently developed Python package ({\bf defringe}) based on the previously used {\sc iraf} defringing routines (\href{https://www.stsci.edu/files/live/sites/www/files/home/hst/instrumentation/stis/documentation/instrument-science-reports/_documents/1997_16.pdf}{Walsh et al. 1997}; \href{https://www.stsci.edu/files/live/sites/www/files/home/hst/instrumentation/stis/documentation/instrument-science-reports/_documents/1998_19.pdf}{Goudfrooij et al. 1998}; \href{https://stistools.readthedocs.io/en/latest/defringe.html}{readthedocs}).
In order to maintain high S/N in the defringed spectral images, a series of 5-6 extra fringe flat exposures, with exposure times significantly longer than the default value, was obtained in the occultation period of each orbit (Tables~\ref{tab:obs} and \ref{tab:obs2}).
Each set of those long fringe flat exposures was processed as usual in {\bf calstis}, then the individual exposures were combined (with cosmic ray correction via {\bf ocrreject}) into a single deep fringe flat image for that orbit.
Each row in the combined fringe flat was then normalized via a spline fit -- which also removed the horizontal striping in the images -- and all pixels with x $<$ 250 were set to 1.0 (to remove the signature of the order sorter fringes at the shortest wavelengths).
Because the Python {\bf defringe} package currently expects full-size (1024$\times$1024) images, each 55 Cnc sub-array scanned image was padded with zeros -- essentially placing it back into position in the full CCD format.
The overall offset (in x) and scaling of the normalized fringe flat for each science exposure was then determined by a cross-correlation procedure -- considering a range in y of 250 pix (centered at y=512), typically stepped in x by 0.1 pix from $-$1.0 to +0.5 and stepped in scale by 0.04 from 0.8 to 1.2; see Appendix Table~\ref{tab:parms}.
The scaled, shifted fringe flat was then applied to the padded science exposure.
While cross correlations of the fringe patterns for individual rows in the full-frame scanned images obtained in orbit 1 of program 15383 suggest that the x-offset between the fringe flat and science exposures may vary by several tenths of a pixel over the full spatial extent of the detector, such differences are small compared to the scale (in x) of variations in the fringing pattern.
Fringe corrections using single, average values of the x-offset and scaling for each image do thus appear to work reasonably well for the sub-array scans (Fig.~\ref{fig:defr}), and the 1D spectra extracted from the fringed and defringed 2D spectral images appear to be quite consistent in flux (Fig.~\ref{fig:extr}).


\ssubsection{1D spectral extraction}\label{sec-extr}

Spectra were extracted directly from the flt files after defringing, with no background subtraction or 2D rectification.
In preparation for the detrending analysis in Section~\ref{sec-detr}, the full extent of the scanned flux was extracted, using an extraction box of 260 pixels.
Because of the relative uniformity of the flux over that large region, the {\bf calstis} {\bf x1d} procedure sometimes had trouble identifying the center of the spectral trace, and so the center was fixed at the central row (i.e., a2center=512, maxsrch=0).
The 260-pixel window was large enough to ensure that essentially all of the flux was extracted, even if the actual center of the scanned region were to differ from the assumed value (512) by a few pixels.
(Image-to-image shifts in the x,y position and wavelength zero point of the scans were typically much smaller -- of order several tenths of a pixel; see Appendix Table~\ref{tab:parms}.)
Spectra obtained with an extraction window of 300 pixels were consistent with those from the smaller extraction window.
While extraction along columns from the un-rectified spectral images introduces a slight smearing in wavelength, that smearing is much smaller than any of the coarse spectral bins examined below.
And while we have chosen to extract all of the flux obtained for each scan into a single spectrum, one could extract multiple, more limited spatial/temporal segments to obtain higher time resolution (which currently reflects the 9-10 scans obtained during each full orbit).

For converting count rates to physical flux, we note that the adopted 260-pixel window does not have a defined aperture correction in the PCTAB reference file.
The nearest windows in the PCTAB are at 200 pixels and 600 pixels, so {\bf calstis} uses the 200 pixel aperture correction -- yielding a slight over-correction in this case -- when converting from count rates to flux units.
The analysis below uses the count rates, to focus on the relative changes in flux over time.
Custom aperture corrections would be needed for corresponding studies of the absolute fluxes.


\ssubsection{Stare-mode observations}\label{sec-stare}

The bleeding of charge along columns in saturated stare-mode observations produces a wavelength-dependent broadening of the spatial profile of the spectrum and also removes spatial information within the saturated columns.
These effects complicate the defringing of the spectral images, the identification and removal of cosmic-rays, and the extraction of the 1D spectra.

\begin{figure}[b!]
\centering
\includegraphics[scale=0.15]{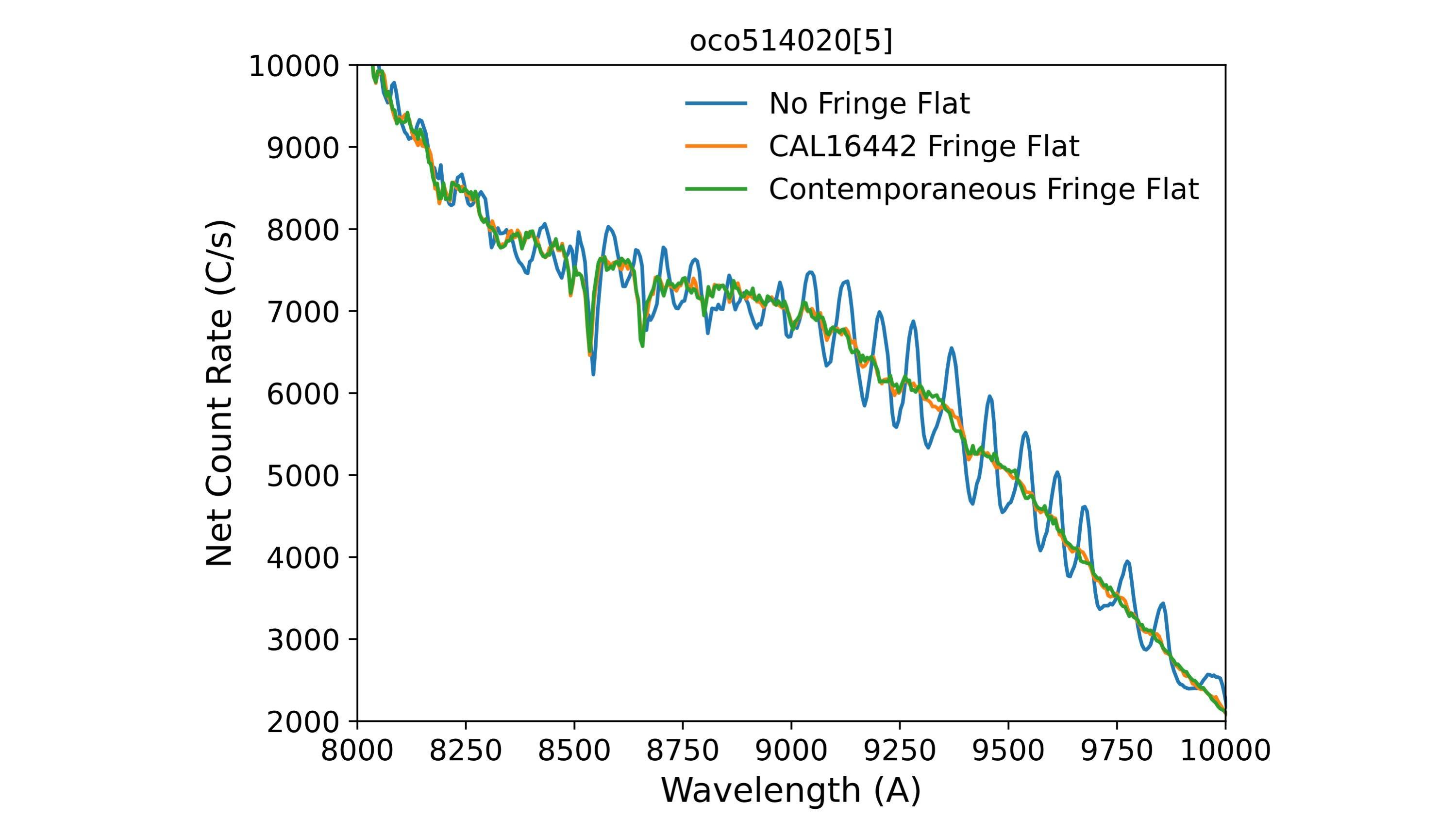}
\caption{{\small 1D spectra from the fifth stare-mode exposure in the second orbit of visit 2 of program 13665.
The blue curve shows the spectrum with no defringing correction.
The fringes are more prominent than those in the uncorrected spectrum in Fig.~\ref{fig:extr} because the counts are summed over a much smaller spatial extent.
The yellow shows defringing done with a full-frame fringe flat taken $\sim$6 years later as part of program 16442.
The green shows defringing done with a contemporanous flat, with only the central 7 pixels of the spectral trace corrected (see text).
In this case, the two defringing corrections yield very similar results.}}
\label{fig:jl1}
\end{figure}

For defringing the stare-mode STIS observations, program 13665 obtained a single contemporaneous fringe flat at the end of each visit through the 0.3x0.09 aperture, as recommended for point-sources in the \href{https://hst-docs.stsci.edu/stisihb/chapter-11-data-taking/11-2-exposure-sequences-and-contemporaneous-calibrations#id-11.2ExposureSequencesandContemporaneousCalibrations-11.2.3FringeFlatFields}{STIS Instrument Handbook} (see also \href{https://www.stsci.edu/files/live/sites/www/files/home/hst/instrumentation/stis/documentation/instrument-science-reports/_documents/1997_15.pdf}{Baum et al. 1998}).
(While using the same slit as the science observations (in this case 52x2) is suggested for diffuse objects, a smaller slit is generally recommended for point source observations in order to better simulate the illuminating lamp as a point source.)
The {\bf stistools defringe} procedure identifies which aperture was used for the fringe flat and creates a normalized, scaled, and shifted fringe flat appropriate for the science observations.
For the 0.3x0.09 aperture, the normalized fringe flat is only calculated for 7 pixels around the trace (i.e., the spatial extent of the aperture and the height of the default extraction aperture for a point source observed with the CCD).
Unfortunately, because the stare-mode observations of 55 Cnc are so highly saturated, the flux in the science observations extends well beyond that 7-pixel extraction aperture.
Attempts to expand the extent over which the normalized fringe flat is defined resulted in increased noise further from the central trace, so we only corrected the central 7 pixels along the trace with the contemporaneous fringe flat.
Nevertheless, use of that default fringe flat does significantly reduce the fringing in the extracted spectra -- largely because the throughput and sensitivity of the detector drop off at the wavelengths where fringing is most problematic.
As a result, the width of the spatial profile is actually not much larger than 7 pixels beyond 8000 \AA\ -- so that the default fringe flat does correct the majority of the flux (see Figure~\ref{fig:jl1}).

We also experimented with using the fringe flats from program 16442, which used the larger 52x0.1 aperture for the fringe flats, to defringe the program 13665 science data.
Despite having been taken $\sim$6 years later, the program 16442 fringe flats do generally yield results consistent with those obtained using the contemporaneous (program 13665) fringe flats (see Figure~\ref{fig:jl1}).
However, in the rest of this work, we use the data defringed using the contemporaneous flats.

As was done for the scanned data, the stare-mode spectra were extracted directly from the flt files after defringing, skipping background subtraction and 2D rectification.
In the stare-mode spectra, the bleeding along columns from the saturation produced spatial profile widths of up to about 25 pixels at the shorter wavelengths -- well-beyond the standard 7-pixel extraction height (Fig.~\ref{fig:rawstare}).
We therefore chose an extraction height of 32 pixels for the stare-mode data, to ensure that all of the flux would be measured.
Unlike with the scanned data, we let the spectral extraction algorithms in {\bf calstis} locate the center of the spectral trace.

The loss of spatial information when charge bleeds along columns in saturated stare-mode data complicates attempts to identify cosmic rays within the spectral profile.
Because of this, the traditional cosmic-ray algorithms described in Section~\ref{sec-cr} will not work.
The cosmic rays in the 55 Cnc stare-mode data were therefore identified solely through the time series in the 1D extracted spectra, rather than through spatial comparisons within the individual 2D spectral images.
After spectral extraction, cosmic rays can be recognized at wavelengths with unexpectedly large flux, compared to the values for other exposures obtained close in time.
Therefore, to identify cosmic rays in the saturated stare-mode images, we compared the flux at each wavelength to the fluxes at that wavelength in the two spectra closest in time (as in \href{https://ui.adsabs.harvard.edu/abs/2018AJ....155...66L/abstract}{Lothringer et al. 2018}).
At each wavelength, if the average difference in flux between the current spectrum and the two nearest spectra is greater than 5$\sigma$, that flux value is replaced with the average from the two nearest spectra.
This procedure identified a few major cosmic-ray hits in these relatively short exposures, without altering any astrophysical features.


\ssection{Detrending analysis}\label{sec-detr}

Previous time series studies of exoplanet host stars with HST -- whether based on pointed/saturated observations with one of the STIS first-order gratings or spatial scans with one of the WFC3 grisms -- have generally found both systematic variations in the fluxes within each orbit of a given visit and overall orbit-to-orbit offsets in the fluxes.
As noted above, the first orbit in a given visit is often discrepant, and the first observation in each subsequent orbit often yields a noticeably different flux from the values found for the rest of the observations in that orbit.
The systematic intra-orbital variations are often ascribed to thermal cycling (``breathing'') of the telescope in its orbit, which can affect the focus.
The discrepant fluxes are thought to be due to the telescope adjusting to the thermal environment of a new pointing or a new visibility period, and are often discarded.
These systematic (apparently) instrumental effects are not well understood, however, as the patterns in the relative fluxes often differ for different objects, different visits for the same object, and different wavelength intervals in a given set of observations.

Functionally, the model for the flux that is used to fit the data follows the form
\begin{center}
$f(t) =  T(t,\theta) \times F_0 \times S(x)$,
\end{center}
where $f(t)$ is the total flux over time ($t$), $T(t,\theta)$ is the theoretical transit model, with $\theta$ representing the transit parameters of the planet (including the planet-to-star radius ratio, the stellar limb darkening, the time of transit center, and the orbital inclination, separation, and period).
For the transit model, we use {\bf batman} (\href{https://ui.adsabs.harvard.edu/abs/2015PASP..127.1161K/abstract}{Kreidberg 2015}).
Finally, $F_0$ is the total baseline flux of the star (usually normalized over), and $S(x)$ represents the systematics model.

Previous studies have often used at least a visit-long linear baseline with time and up to a fourth-order polynomial describing orbit-long systematics as a function of HST’s orbital phase, $\phi_{\rm HST}$:
\begin{center}
$S(x) = p_1 t + p_2 \phi_{\rm HST} + p_3 \phi_{\rm HST}^2 + p_4 \phi_{\rm HST}^3 + p_5 \phi_{\rm HST}^4$,
\end{center}
where $p_i$ represents the coefficient fit to the respective vector’s trend with flux.
While this parameterization can generally fit most of the systematic trends, there are often residuals that can be fitted with other information.
For the visits including a transit of 55 Cnc e, we used a form of ``jitter detrending'' (\href{https://ui.adsabs.harvard.edu/abs/2019AJ....158...91S/abstract}{Sing et al. 2019}),  where header information and engineering data from the ``jitter'' (.jit) files are used as vectors to detrend against, with polynomial coefficients fitted to the data.
More detailed information regarding trends in the parameters included in the jitter files, for the observations obtained in programs 15383 and 16442, is given in Appendix B.
 
In addition to time and orbital phase, we tried a variety of other vectors.
These included the right ascension and declination of the aperture reference, the position angle, the angle between the aperture and the Moon or Sun, the altitude of the Sun above the horizon, the CCD housing temperature, the FGS V2 and V3 dominant and roll coordinates, and the angle between the HST zenith and the target.
After trial and error, both the declination of the aperture reference and the angle between HST zenith and the target were found to correlate with the flux such that they could be used for detrending.
We used fits to these vectors up to second order in the systematics model for both the white-light and spectroscopic bins and for both the scan-mode and stare-mode data.

Each visit was analyzed independently.
In each case, we neglected the first orbit of data as well as the first exposure of each orbit as those have historically exhibited greater systematic noise.
For the visits that included a transit of 55 Cnc e, we first performed fits to the white-light curve to check the transit timing -- fixing the orbital inclination, separation, and period to the parameters from the ExoClock project in \href{https://ui.adsabs.harvard.edu/abs/2023ApJS..265....4K/abstract}{Kokori et al. (2023)}.
We performed fits both fitting the transit center time (T$_0$) and fixing it to the ephemeris from \href{https://ui.adsabs.harvard.edu/abs/2023ApJS..265....4K/abstract}{Kokori et al. (2023)}. 
For the scan-mode visit and the first and last stare-mode visits, the fitted T$_0$ was consistent with the published ephemeris. 
Visit 2 of the stare-mode data, however, preferred a fit to T$_0$ that differed by about 25 minutes compared to the ephemeris. 
That discrepancy did not change by fitting different systematics models, and it appears in spectroscopic bins as well. 
We therefore present analyses for the spectroscopic bins that have T$_0$ both fixed and free. 
If T$_0$ was free, we fixed it for the spectroscopic light curves.

For both the white-light curve and the spectroscopic light curves, we used quadratic limb-darkening coefficients calculated using the Exoplanet Characterization Toolkit’s limb-darkening calculator (\href{https://ui.adsabs.harvard.edu/abs/2021zndo...4556063B/abstract}{Bourque et al. 2021}) with PHOENIX models (\href{https://ui.adsabs.harvard.edu/abs/2013A&A...553....6H/abstract}{Husser et al. 2013}), using stellar parameters from \href{https://ui.adsabs.harvard.edu/abs/2018A&A...619....1B/abstract}{Bourrier et al. (2018)}.
The limb-darkening coefficients were fixed in all analyses. 
The only other fitted parameters of the white-light curve were the transit radius ratio (Rp/Rs) and the coefficients of the systematics model.
We used a Levenberg-Marquardt method to fit the data and to compute uncertainties.


\ssection{Results}\label{sec-res}

\begin{figure}[b!]
\centering
\begin{minipage}[c]{0.5\textwidth}
   \includegraphics[scale=0.35]{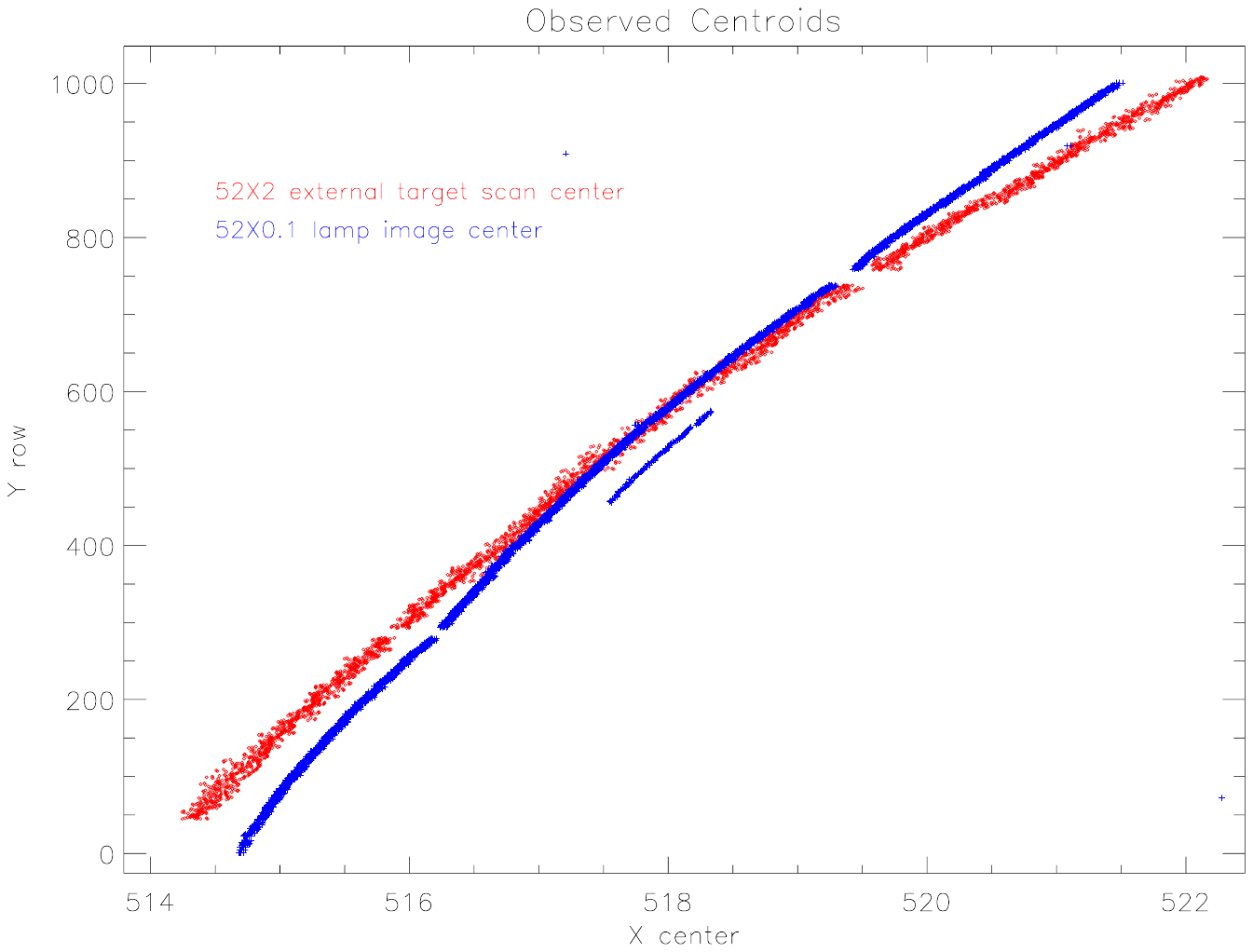}
\end{minipage} \hfill
\begin{minipage}[c]{0.4\textwidth}
   \caption{{\small Observed centroids of images of GRW+70~5824 (scanned along the 52x2 aperture; red) and of the tungsten lamp (through the 52x0.1 aperture; blue), from visit 1 of program 15383.
   The difference in angle between the two, estimated from linear fits, is about 0.065 degrees.
   While it is not yet known whether that difference is typical, the scans in visit 2 of the program were adjusted accordingly.}}
   \label{fig:geom}
\end{minipage}
\end{figure}

\ssubsection{Scan geometry and positional stability}\label{sec-geom}

The imaging scans of GRW+70~5824 taken during the first visit of program 15383 were designed to compare the alignment of spatial scans performed along a wide slit (52x2) with flat field lamp exposures taken through a narrower slit (52x0.1). 
Using the wide 52x2 slit for science targets is desirable both to minimize flux losses during the exposures and to enable the most accurate flux calibration of the spectra.
Using the narrow 52x0.1 aperture for the flat field is desirable to simulate (as much as possible) the illumination from a trailed point source, in order to obtain a similar fringing pattern.
In order to obtain the most effective defringing of spectra taken in scan mode, those two images should be aligned as closely as possible.

Examination and analysis of those images provided several important pieces of information that were then used in planning the spectral observations obtained in the second visit of the program:
\begin{itemize}
\item{The starting point for the forward scans was quite repeatable (to within several tenths of a pixel), but the starting point for the corresponding reverse scans in the round-trip mode exhibited a scan-rate-dependent offset to higher than expected y values, most likely due to a timing issue. 
While round-trip scans have been used successfully for WFC3 spatial scans, it apparently required considerable experimentation and adjustment of the commanding to obtain the desired round-trip behavior.
Until such an effort can be made for STIS round-trip scans, it is recommended to perform only forward scans with STIS.}
\item{Using the nominal scan angle of 90 degrees, the resulting trailed stellar image was tilted slightly, by $\sim$0.065 degrees, relative to the image of the lamp (Fig.~\ref{fig:geom}).
While it is not known whether that slight difference is typical, all of the scans in the second visit of the program used a scan angle of 90.065 degrees.  
For that scan angle, cross-correlations of the spectral images with the corresponding contemporaneous flat-field images indicated good alignment of the fringing (in x, as a function of y on the CCD) -- within a few tenths of a pixel over the extent of the sub-array scans (much smaller than the scale of the fringing variations).}
\end{itemize}

Spatial offsets among the scanned spectral images and the flat-field images obtained in visit 2 of program 15383 were determined both via cross-correlations of normalized rows and columns in the raw and flt images (i.e., before removal of cosmic rays and fringing) and in the defringing.
Examination of those offsets (in both x and y) provides some information regarding the positional reproducibility and alignment of the various images:
\begin{itemize}
\item{The starting and ending points (in y) of the exposed section of the scanned spectra -- as delimited by the ``ramps'' up to and down from the roughly uniformly exposed plateau region of the scans -- were quite consistent, differing by at most several tenths of a pixel over the course of each orbit.}
\item{The fringing in the deep flat-field exposures is also quite consistent, with mean offsets in the fringing pattern less than 0.1 pixel in both x and y (though there are slight changes with y for both).
As such offsets are much smaller than the typical scale of the variations in the fringing pattern, it was thus straightforward to combine the individual flats obtained during each occultation to produce a single, high-S/N fringe flat for each orbit.}
\item{Somewhat larger x-offsets were seen for the fringing in the trailed stellar exposures, compared to the combined deep fringe-flat exposures -- with mean x-offsets of 0.1-0.6 pixels and tendencies to decrease slightly with y and to increase slightly (by $\sim$0.3 pix) over the course of each orbit (Appendix Table~\ref{tab:parms}).}
\item{The x-offsets determined in the defringing appear to be correlated with both the orbital phase, the phase-dependent focus, the x-offsets determined by independent cross-correlations of image rows, and the offsets in the dispersion direction obtained in the detrending (Appendix Table~\ref{tab:parms}).}
\item{In the defringing, the scale factors for the fringe flats are all $\sim$0.9.} 
\end{itemize}

\begin{figure}
\centering
\includegraphics[scale=0.7]{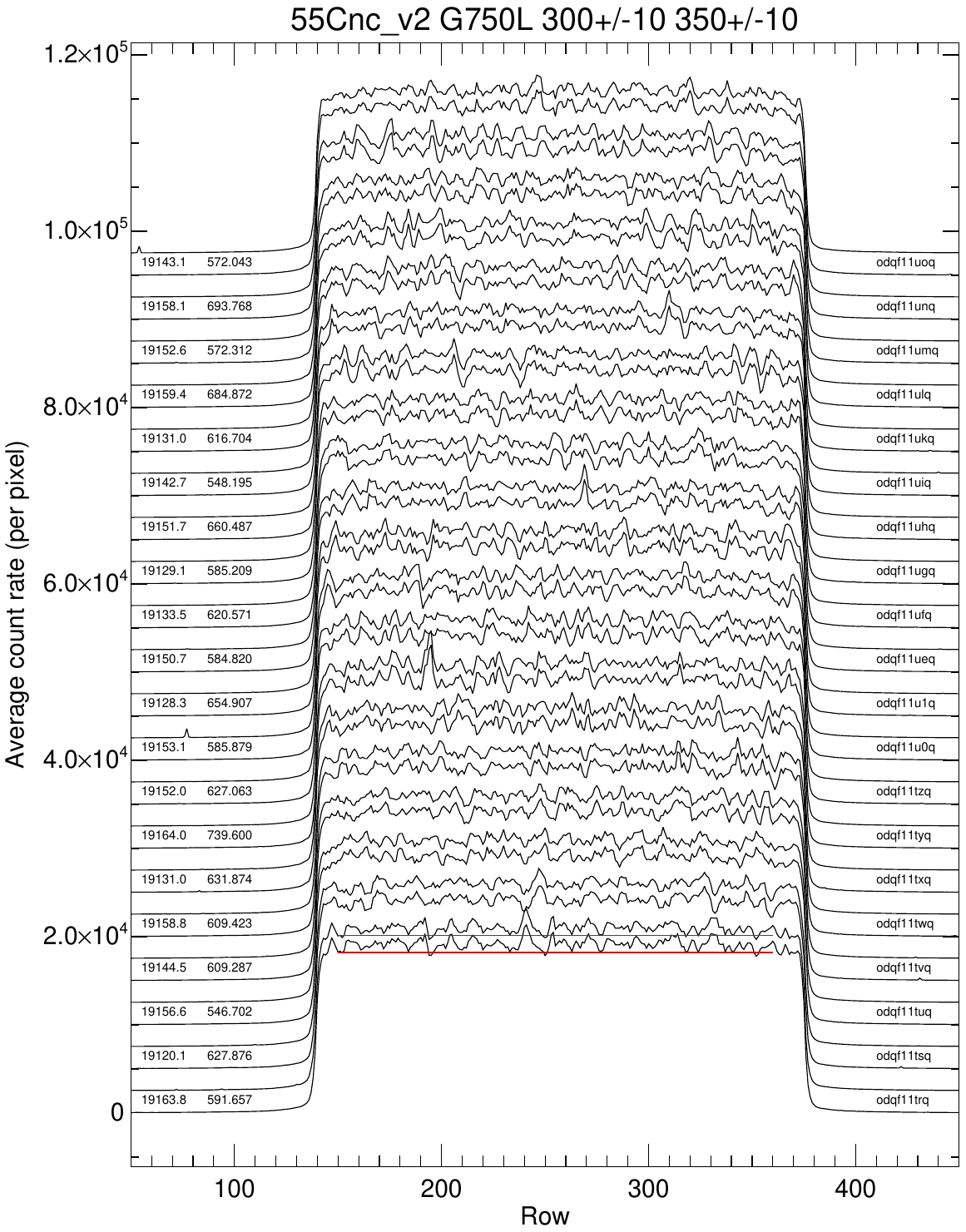}
\caption{{\small Average count rates (per pixel) along columns 290-310 and 340-360 (lower and upper curves for each exposure), for the 20 scanned exposures obtained in visit 2 of program 15383.
The average and standard deviation of the count rates in the plateau region are given on the left for each exposure; the red horizontal lines bracketing the lowest curve mark $\pm$5\%.
The fluctuations seen for the two column ranges in each exposure are very similar, but the specific patterns in those fluctuations differ from exposure to exposure.}} 
\label{fig:stripe}
\end{figure}

\ssubsection{Flux stability}\label{sec-flux}

As one of the motivations for performing spatial scans has been to obtain time series observations of stellar flux in either broad or narrow wavelength regions, a number of comparisons were made to assess the consistency of the count rates within individual scans, between different scans taken in the same orbit, and between scans taken in different orbits within the same visit.
The most varied and detailed comparisons were made for the 20 sub-array scans obtained over orbits 2 and 3 of the second visit of program 15383, for which the stellar flux was expected to be essentially constant.

Apparent variations in the count rate during individual scans -- as evidenced by the roughly horizontal ``striping'' seen in the 2D images (e.g., Figs.~\ref{fig:rawfull}, \ref{fig:fltsub}, and \ref{fig:defr}) -- have already been noted.
Figure~\ref{fig:stripe} compares the average count rates (per pixel) along columns 290-310 and 340-360, corresponding to relatively narrow wavelength intervals outside the region affected by fringing, for the 20 scanned exposures (flt files) from program 15383.
The average and standard deviation of the count rates in the plateau regions are given on the left; the typical scatter is $\sim$3\%, with maximum deviations $\sim$10\%.
For the 20 exposures, the scatter in the average plateau count rates is less than 0.1\%.
For each exposure, the patterns seen in the two column intervals are very similar, indicating that the observed fluctuations are not due to noise.
While some of the striping may be common among multiple images, most of it appears to be specific to the individual images -- perhaps due to slight variations in the scan rate (which was $\sim$1.1 pix per second) during the individual scans.
Initial attempts to compare the striping with the ``jitter'' seen in more finely sampled (in time) versions of the associated jitter files -- projecting the jitter along the scan path -- were inconclusive, however.
Some of the striping features in the scanned spectra seem similar to the ``flicker'' seen in some WFC3 spatial scans, but the origin of those variations in the WFC3 data appears to be specific to the WFC3 (S. Baggett and O. Lupie, private communications, 2022). 
The presence of this striping in the STIS spatial scans may complicate attempts to achieve finer time resolution by sub-sampling the individual scans -- but that has not yet been investigated in any detail.

\begin{figure}[b!]
\centering
\includegraphics[scale=0.7]{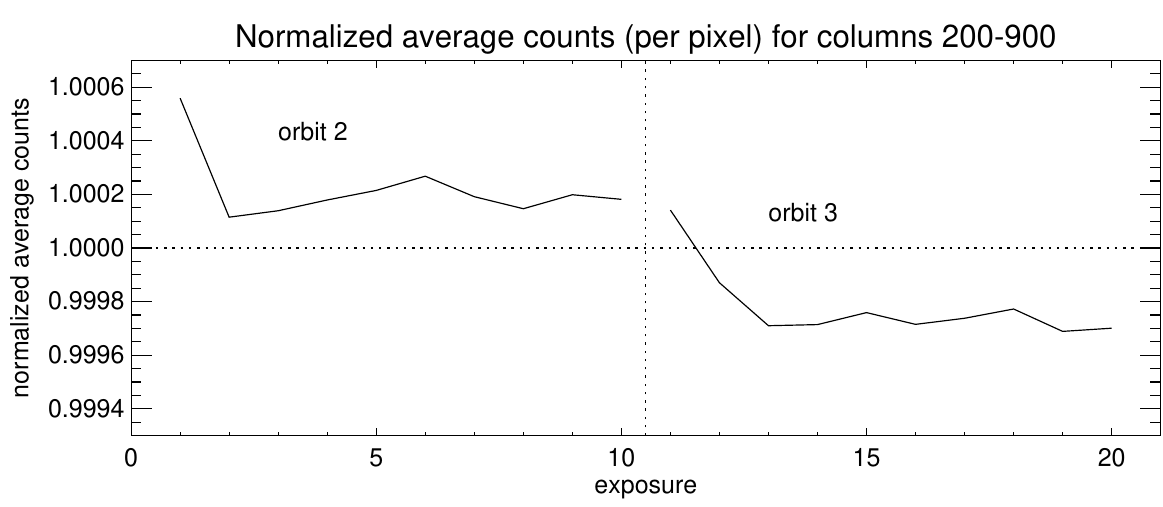}
\caption{{\small Normalized average counts (integrated over about 3500 \AA, or $\sim$70\% of the wavelength coverage) for the individual short scans taken in orbits 2 and 3 of program 15838, visit 2.
(Note that these are for the flt files, which are not corrected for cosmic rays or fringing.)
Within each orbit, the fluxes agree to within $\sim$0.01\% (100 ppm) -- and for the last 7-8 scans in each orbit they agree to within 0.005\% (50 ppm).
While the variations within each orbit are very similar, there appears to be an overall offset of about 0.04\% between the two orbits.}}
\label{fig:flux}
\end{figure}
\begin{figure}
\centering
\includegraphics[scale=0.7]{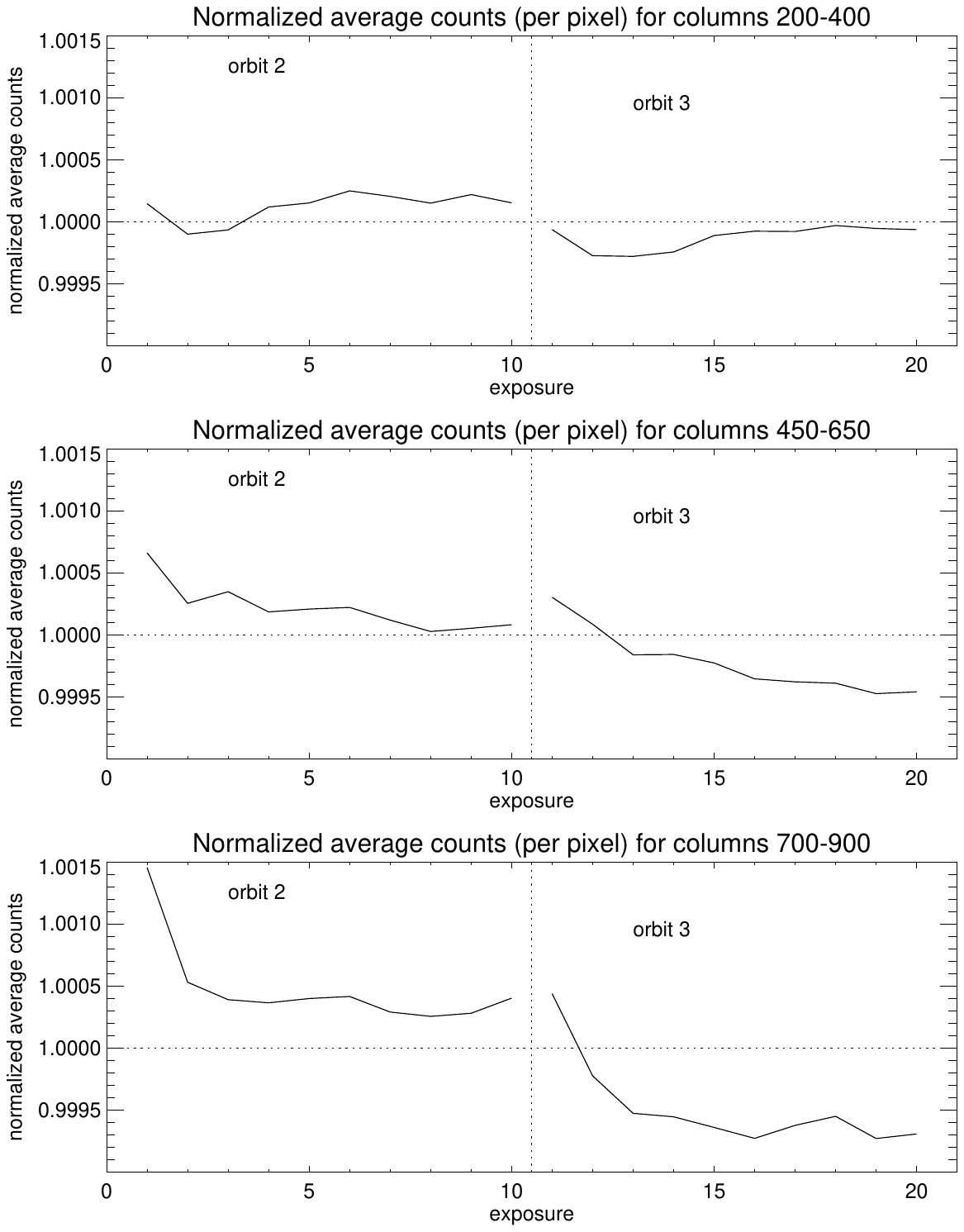}
\caption{{\small Normalized average counts (integrated over three 200-pix ranges, each $\sim$20\% of the full wavelength coverage) for the individual short scans taken in orbits 2 and 3 of program 15838, visit 2.
(Note that these are for the flt files, which are not corrected for cosmic rays or fringing.)
The three wavelength segments exhibit somewhat different trends within each orbit, with slightly lower count rates for all three in orbit 3.}}
\label{fig:flux3}
\end{figure}

Expanding the range of columns to encompass most of the covered spectral range -- and summing over all the included pixels -- provides a measure of the total ``white-light'' flux obtained in the observations.
For those overall sums, the most reliable results were obtained by including the ``ramp'' regions in each column (in addition to the more uniformly exposed plateau regions) -- covering a span of $\sim$260 pixels (in y) for the sub-array scans obtained in program 15383.
The total white-light count rates for the 20 exposures (for columns 200-900 in the flt files), normalized to the average value for the 20 exposures in that series, are shown in Figure~\ref{fig:flux}.
Within each orbit, the total count rates are highest for the first exposure, but settle down to a nearly constant value for the last 7-8 exposures, with scatter of order 0.005\% (50 ppm).
There is a slight overall offset in the normalized count rates between the two orbits, of order 0.04\% (400 ppm), but the general trends in the variations within each orbit are very similar (and remain so after correcting for cosmic rays and fringing).
The intra-orbital trends in the relative count rates can exhibit some differences, however-- for different visits (e.g., program 15383 vs. program 16442) and for different wavelength ranges within a given visit (Figure~\ref{fig:flux3}); see also the discussions and associated figures in the following sections. 


\ssubsection{Detrending fits}\label{sec-detrfits}

As the scanned spectra obtained in visit 2 of program 15383 covered a brief, roughly 2.4-hr period in which 55 Cnc e was out of transit, the stellar flux was expected to be fairly constant for all the exposures (though see \href{https://ui.adsabs.harvard.edu/abs/2019A&A...631..129S/abstract}{Sulis et al. 2019}).
That series of exposures thus could provide a useful sample for investigating possible systematic instrumental/observational effects via detrending, if the observed fluxes show any systematic variations -- as seen for the white-light count rates in Fig.~\ref{fig:flux}.
A number of tests of the detrending procedure were performed on those 20 exposures -- varying the detrending parameters, the wavelength range considered, and whether or not defringing was performed.  

\begin{figure}[b!]
\centering
\begin{minipage}[c]{0.3\textwidth}
   \includegraphics[scale=0.55]{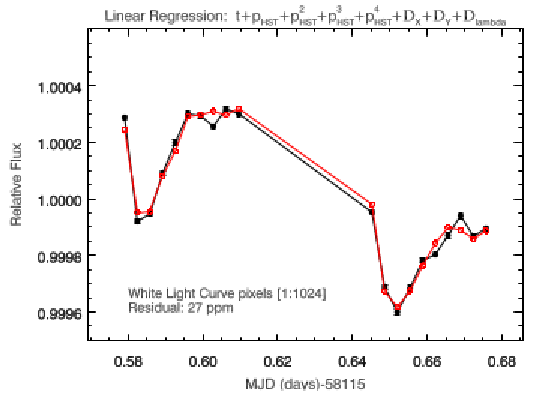}
\end{minipage} \hfill
\begin{minipage}[c]{0.3\textwidth}
   \includegraphics[scale=0.55]{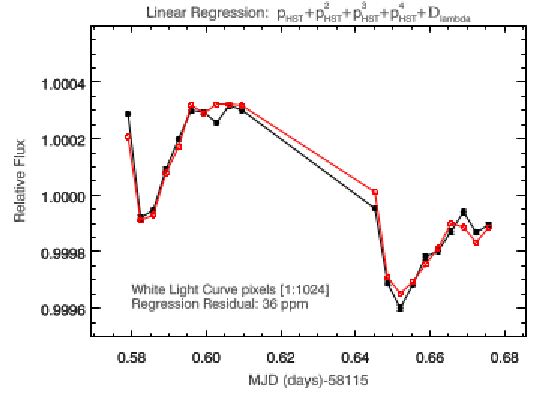}
\end{minipage} \hfill
\begin{minipage}[c]{0.3\textwidth}
   \includegraphics[scale=0.55]{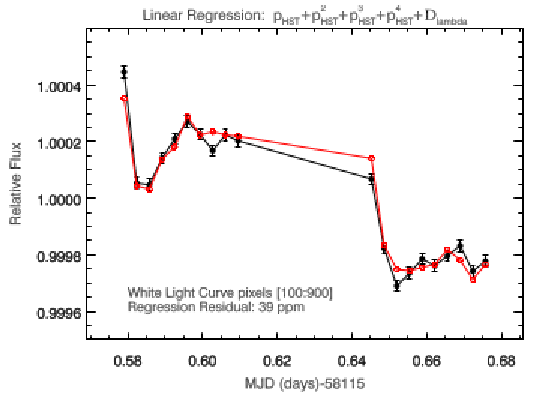}
\end{minipage}
\caption{{\small Linear regression (detrending) fits to fluxes derived from scanned spectra (program 15383, visit 2), for slightly different parameterizations and/or wavelength ranges.
(Note that these are for the flt files, which are not corrected for cosmic rays or fringing.)
({\it left}) White light curve (full wavelength range), fitted with a model parameterized by time, HST phase, X, Y, and D.
({\it center}) White light curve fitted with a model parameterized by HST phase and D.
({\it right}) White light curve (for a slightly restricted wavelength range) fitted with a model parameterized by HST phase and D.
The residuals for the three fits are 27, 36, and 39 ppm, respectively.}}
\label{fig:detrfits}
\end{figure}

Figure~\ref{fig:detrfits} shows the results of several such tests performed on the flt files from program 15383, over fairly broad spectral ranges.
In the left-hand panel, the full spectral range is included, with a fit parameterized by time, HST phase (up to 4th-order terms), and the displacements in x, y, and dispersion.
In the center panel, the time and the x and y displacements were dropped.
In the right-hand panel, the same parameters as in the center were used, but the wavelength range was restricted by dropping the lowest and highest $\sim$100 pixels (in x).
Changing the wavelength range produced slight changes in the intra-orbit trends (and see also Fig.~\ref{fig:flux}).
The residuals remaining after detrending are 27, 36, and 39 ppm, respectively -- comparable to the best results from previous HST observations.

\begin{figure}[t!]
\centering
\includegraphics[scale=0.15, clip = true, trim = 0 200 0 200]{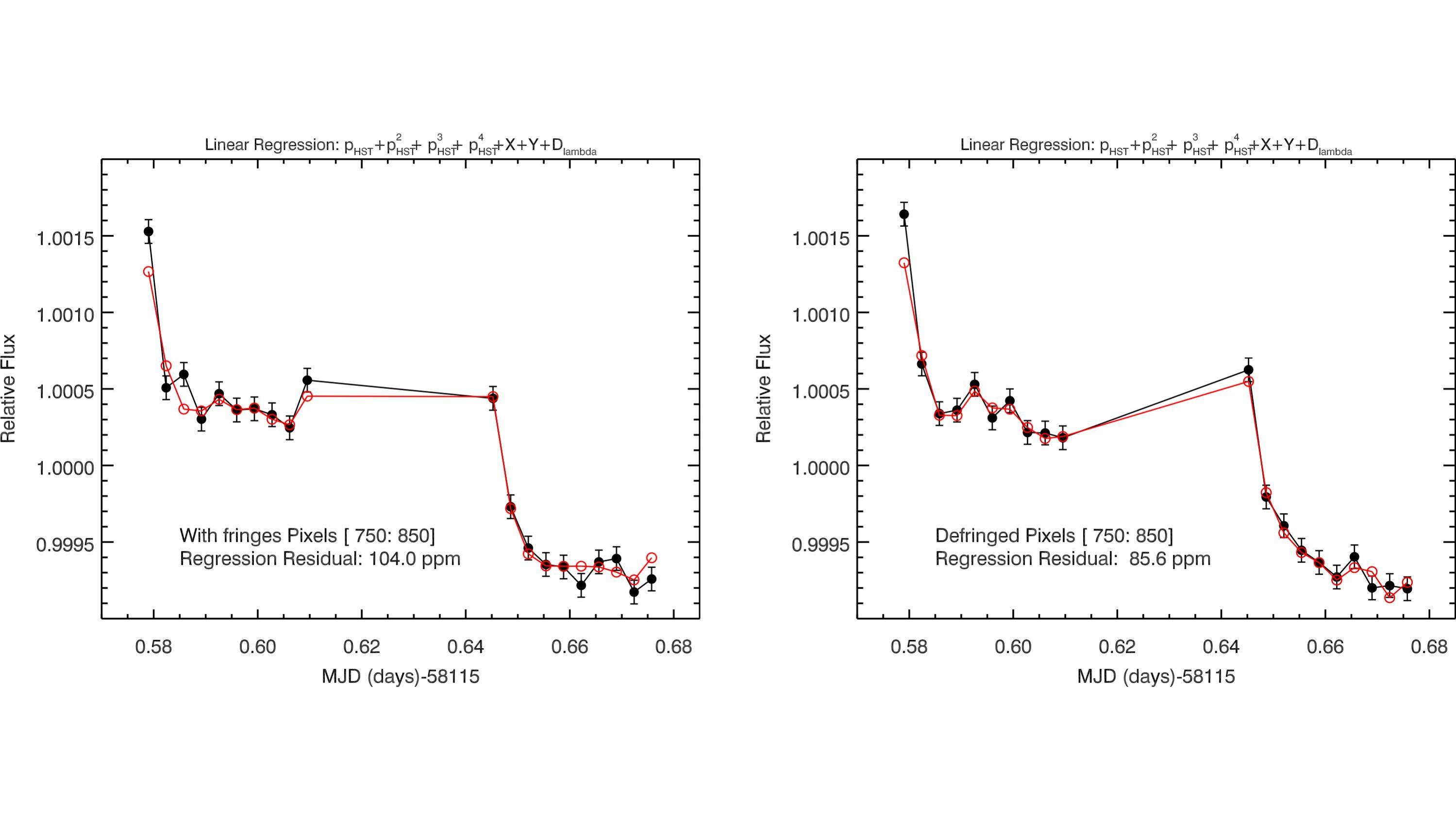}
\caption{{\small  Linear regression fits to fluxes for a wavelength interval near the red end of the coverage (program 15383, visit 2) -- before defringing ({\it left}) and after defringing ({\it right}).
The residuals are higher than for the fits to the white-light curves in Fig.~\ref{fig:detrfits}, due to the restricted wavelength range and lower count rates, but (over this smaller range) are smaller by 15-20\% for the defringed data.}}
\label{fig:detrdefr}
\end{figure}

\begin{figure}[b!]
\centering
\includegraphics[scale=0.15, clip = true, trim = 0 200 0 200]{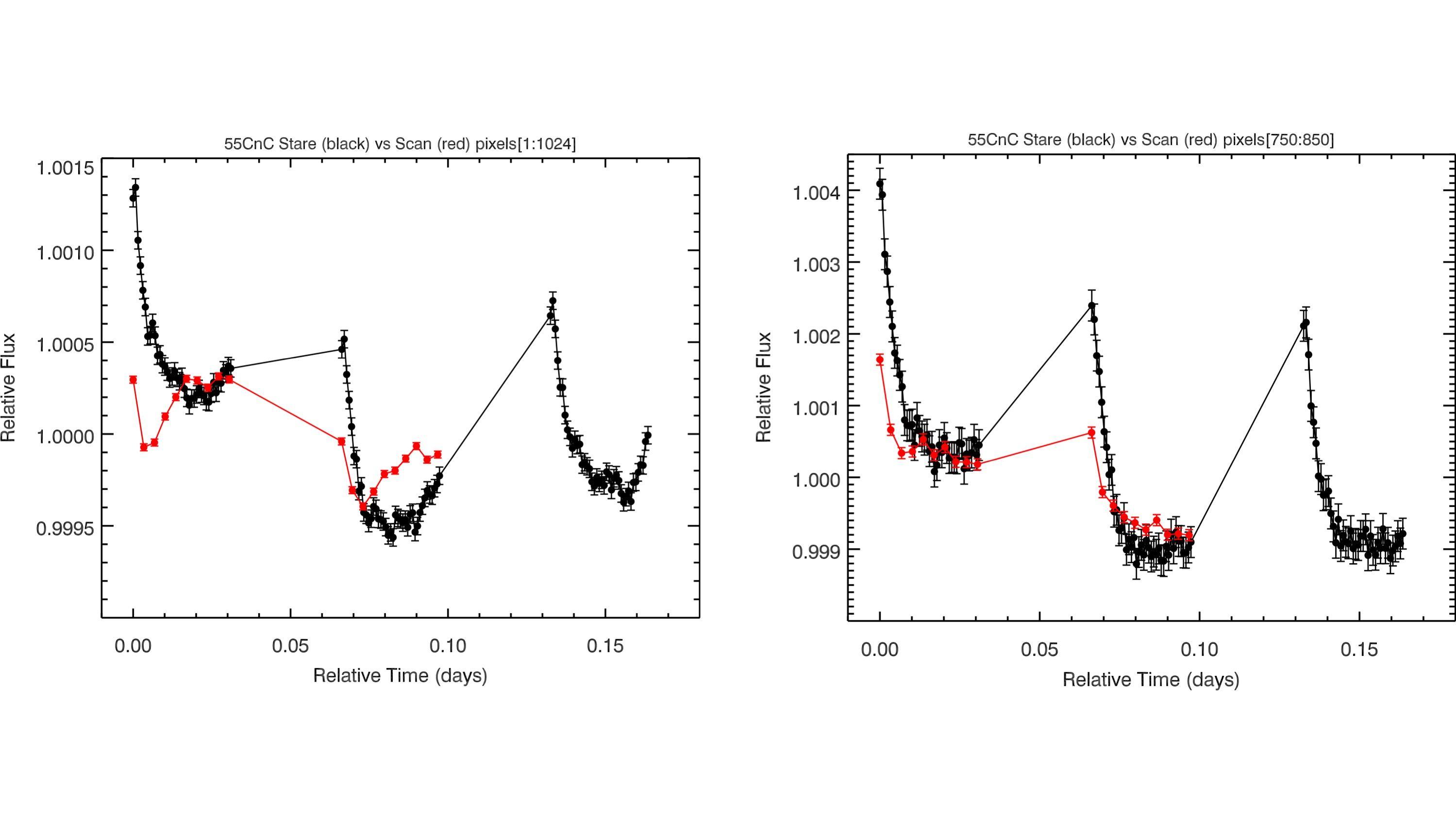}
\caption{{\small Relative fluxes of 55 Cnc – derived from pointed/saturated (stare-mode) STIS spectra (black; program 13665) and from spatially scanned spectra (red; program 15383).
({\it left}) White light fluxes; ({\it right}) fluxes for a smaller wavelength interval toward the red end of the coverage (as in Fig.~\ref{fig:detrdefr}).
The spatially scanned spectra exhibit smaller systematic differences, both from orbit to orbit and within each orbit -- though the latter may be due (at least in part) to the coarser time sampling of the scan-mode data.
(Note also that the second (third) orbit for program 13665 included a transit of 55 Cnc e, while both orbits for program 15383 were out of transit.)}}
\label{fig:scanstare}
\end{figure}

Figure~\ref{fig:detrdefr} compares the detrending fits for a wavelength interval affected by fringing (pixels 750-850), before (left-hand panel) and after (right-hand panel) defringing was performed.
The trends in the relative flux are very similar to those seen for other wavelength ranges, but there are some differences in detail (Figs.~\ref{fig:flux} and \ref{fig:detrfits}).
The detrending fits used the HST phase (up to 4th order) and the x, y, and dispersion displacements.
In this particular case, the residuals from the detrending fits (104 and 86 ppm, respectively) are larger than those found for broader wavelength ranges, but the defringing reduced the residuals by $\sim$15-20\%.

While a detailed discussion of the detrending results for the scanned spectra obtained for program 16442 and for the archival stare-mode data from program 13665 -- all of which included a transit of 55 Cnc e -- is given in the next section (Sec.~\ref{sec-trspec}), some intial comparisons between the results from programs 13665 and 15383 were explored.
Figure~\ref{fig:scanstare}, for example, compares the full white-light curves (left-hand panel) and the curves from a narrower spectral interval toward the red end of the coverage, for the scanned spectra from program 15383 (in red; as in Fig.~\ref{fig:detrfits}) and for the stare-mode spectra from one of the visits from program 13665 (black; including a transit in the second orbit).
Both the scan-mode and stare-mode fluxes exhibit similar intra-orbit trends -- particularly at the longer wavelengths -- and both exhibit overall declines over the extent of the observations.
While the scanned spectra appear to be characterized by smaller systematic differences and scatter, that may be due (at least in part) to the coarser time sampling of the scanned data.

The general similarity in the trends for different orbits in the same visit would seem to suggest that the variations depend largely on the HST orbital phase -- or on some other quantity that depends on or varies somewhat regularly with the phase.
The differences in behavior seen for different wavelength intervals (Fig.~\ref{fig:flux3}), and its non-monotonicity in some cases, however, suggest that the behavior is more complex.


\ssubsection{The optical transit spectrum of 55 Cnc e}\label{sec-trspec}

The transit depths in the spectroscopic light curves form the planetary transmission spectrum, measuring the effective “size” of the atmosphere as a function of wavelength.
This transmission spectrum can be used to identify spectral features coming from the atmosphere, including atomic and molecular absorption or scattering and absorption from aerosols.
In the G750L bandpass, potential opacity sources to expect include the Na and K resonance doublets near 5892/5898 and 7667/7701 \AA, respectively, as well as scattering from aerosols (i.e., small haze or cloud particles) that can cause a slope of larger transit depths towards shorter wavelengths (\href{https://ui.adsabs.harvard.edu/abs/2008A&A...481...83L/abstract}{Lecavelier des Etangs et al. 2008}).

Active regions on the host star’s surface, including star spots and stellar plages, can also induce signals in the transit spectrum.
If the transit chord that the planet follows does not represent the average unocculted surface, then the differential flux that is blocked out can induce redward slopes for star spots (\href{https://ui.adsabs.harvard.edu/abs/2014ApJ...791...55M/abstract}{McCullough et al. 2014}) or blueward slopes for stellar plages/faculae (\href{https://ui.adsabs.harvard.edu/abs/2014A&A...568...99O/abstract}{Oshagh et al. 2014}).

\begin{figure}[t!]
\centering
\includegraphics[scale=0.15]{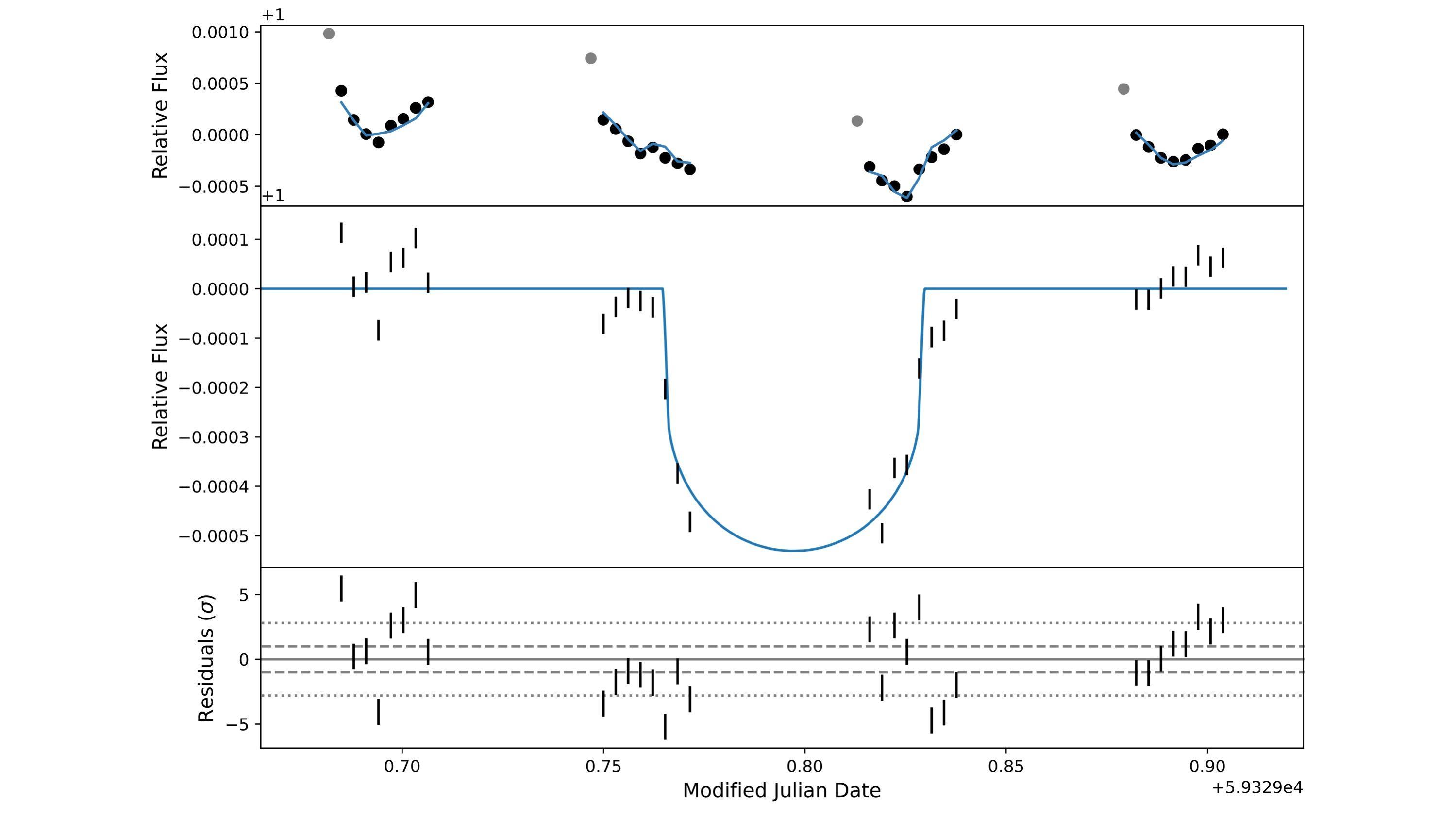}
\caption{{\small White light curve of the spatially scanned data from orbits 2-5 of program 16442.
({\it top}) Raw data. The first exposure of each orbit, indicated in grey, is not included in the fit.
({\it middle}) Systematics subtracted data with the best-fit transit light curve model.
({\it bottom}) Residuals between the model fit and the data.
The transit of 55 Cnc e, with depth $\sim$450 ppm, began during orbit 3 and ended during orbit 4.}}
\label{fig:jl2}
\end{figure}
\begin{figure}[b!]
\centering
\includegraphics[scale=0.14, clip = true, trim = 0 100 0 150]{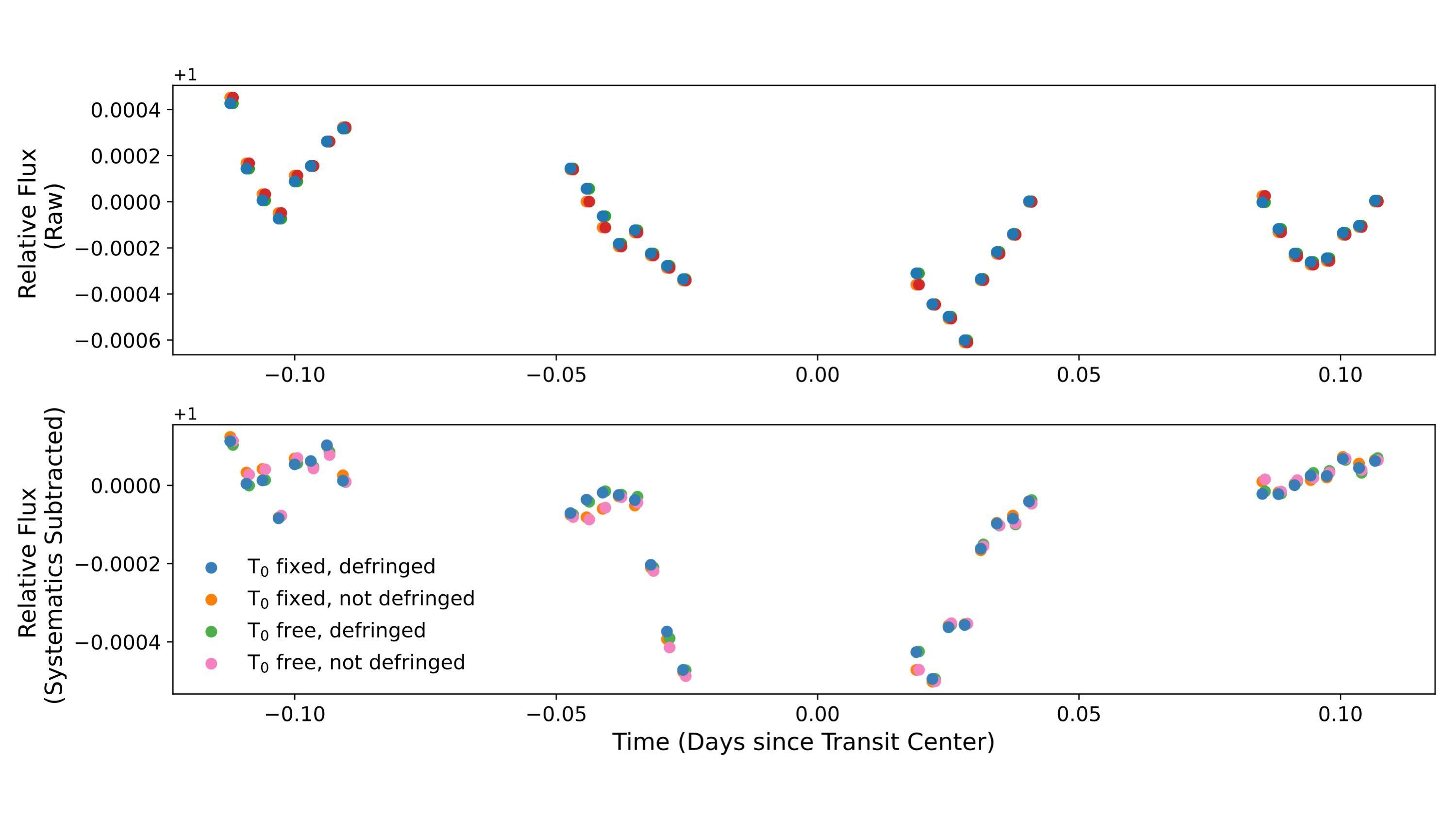}
\caption{{\small White-light light curves of the spatially scanned data from orbits 2-5 of program 16442, for different defringing and transit center scenarios.
({\it top}) Raw data.
({\it bottom}) Systematics subtracted data.
For most of the points, the differences are fairly minor.}}
\label{fig:jl3}
\end{figure}
\begin{figure}
\centering
\includegraphics[scale=0.28]{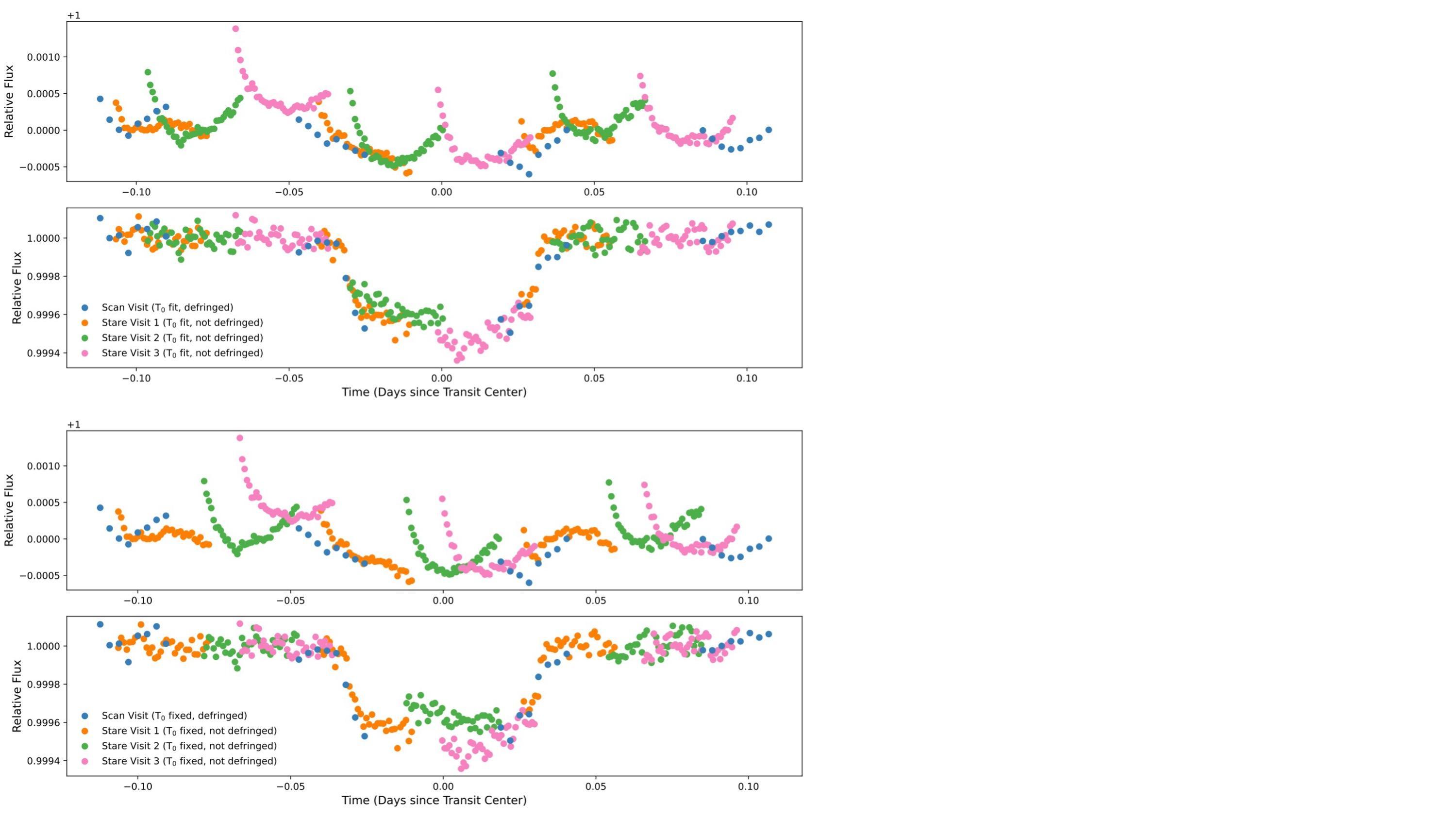}
\caption{{\small ({\it upper two panels}) Raw light curves and systematics detrended light curves for all visits from programs 16442 and 13665, with the transit center ($T_0$) fitted.
The defringed data are shown here for the scan-mode visit, but the non-defringed data are shown for the stare-mode visits (due to the small aperture used for the contemporaneous stare-mode fringe flats).
({\it lower two panels}) Raw light curves and systematics detrended light curves for all visits from programs 16442 and 13665, but with the transit center fixed to the published ephemeris of \href{https://ui.adsabs.harvard.edu/abs/2023ApJS..265....4K/abstract}{Kokori et al. (2023)}.
All visits except visit 2 of the stare-mode data (in green) are consistent with the fitted T$_0$ results.
Note also the apparent differences in transit depth among the visits.}}
\label{fig:jl67}
\end{figure}

\begin{deluxetable}{rcrccccc}
\tablecolumns{8}
\tabletypesize{\footnotesize}
\tablecaption{Detrending results \label{tab:detr}}
\tablewidth{0pt}

\tablehead{
\multicolumn{1}{c}{Dataset}&
\multicolumn{1}{c}{$T_0$\tablenotemark{a}}&
\multicolumn{1}{c}{diff\tablenotemark{a}}&
\multicolumn{2}{c}{scatter (ppm)\tablenotemark{b}}&
\multicolumn{1}{c}{T depth\tablenotemark{c}}&
\multicolumn{2}{c}{Rp/Rs\tablenotemark{d}}\\
\multicolumn{1}{c}{ }&
\multicolumn{1}{c}{(MJD)}&
\multicolumn{1}{c}{(min)}&
\multicolumn{1}{c}{exp}&
\multicolumn{1}{c}{obs}&
\multicolumn{1}{c}{(ppm)}&
\multicolumn{1}{c}{value}&
\multicolumn{1}{c}{err}}

\startdata
16442 v1& 59329.79722 &\nodata & 58 & 45 & 448 & 0.02117 & 0.00027 \\
    scan& 59329.79672 &$-$0.73 & 52 & 37 & 453 & 0.02129 & 0.00027 \\
13665 v1& 56961.06458 &\nodata & 38 & 24 & 392 & 0.01981 & 0.00033 \\
   stare& 56961.06484 &   0.38 & 37 & 25 & 395 & 0.01988 & 0.00033 \\
13665 v2& 57152.56667 &\nodata & 49 & 38 & 316 & 0.01779 & 0.00026 \\
   stare& 57152.58458 &  25.79 & 42 & 29 & 344 & 0.01855 & 0.00031 \\
13665 v3& 57164.35139 &\nodata & 44 & 32 & 466 & 0.02158 & 0.00022 \\
   stare& 57164.35225 &   1.25 & 44 & 30 & 460 & 0.02144 & 0.00028 \\
\enddata
\tablenotetext{a}{Transit center (T$_0$) -- fixed to the \href{https://ui.adsabs.harvard.edu/abs/2023ApJS..265....4K/abstract}{Kokori et al. (2023)} ephemeris value for the first line of each pair, varied for the second line (with difference from the ephemeris value in min).}
\tablenotetext{b}{Rms scatter (data vs. detrending fit) -- expected vs. observed -- in ppm.}
\tablenotetext{c}{Transit depth in ppm.}
\tablenotetext{d}{Transit radius = (transit depth)$^{1/2}$ -- value and error.}
\end{deluxetable}

Figure~\ref{fig:jl2} shows a fit to the scan-mode white-light fluxes from orbits 2-5 of program 16442.
After removing the systematic trends, the transit becomes apparent, beginning near the end of the third orbit (which catches the ingress) and continuing throughout most of the fourth orbit (which catches the egress).
The transit depth, of order 450 ppm, is broadly consistent with previous estimates based on optical spectra (e.g., \href{https://ui.adsabs.harvard.edu/abs/2019A&A...631..129S/abstract}{Sulis et al. 2019}).
Figure~\ref{fig:jl3} illustrates the effects of fixing or fitting the transit center T$_0$ and of defringing the spectra (or not).
For most of the points, those choices had very little effect on the resulting detrended light curve.

Figure~\ref{fig:jl67} shows the raw and detrended time series white-light curves for all four STIS G750L visits including a transit of 55 Cnc e (from programs 13665 and 16442).
The upper and lower panels show the curves with the fitted and fixed transit center times, respectively.
The raw data show large, but repeatable systematic trends, greater in magnitude than the transit signal by a factor of about 3.
The scan-mode curve is generally consistent with the three stare-mode curves, with perhaps a more muted systematics trend -- which could be due to averaging over more pixels on the detector during the scan and/or to the longer exposure times (averaging the systematics over time).
Visit 3 of the stare-mode data shows a larger transit depth, compared to visits 1 and 2.
The scan-mode data appear to resemble more visit 3 of the stare-mode data at the beginning of egress.

Some of the quantities used or derived in the detrending fits to the white-light fluxes (not defringed) from the four visits are given in Table~\ref{tab:detr}. 
For each visit, the first line gives values for the fits with the transit center T$_0$ fixed to the \href{https://ui.adsabs.harvard.edu/abs/2023ApJS..265....4K/abstract}{Kokori et al. (2023)} ephemeris value, while the second line gives the values for the fits with T$_0$ allowed to vary.
All of the fits included the time, HST phase (to 4th order), the declination (to 2nd order), and line of sight zenith angle (to 2nd order).
The rms residuals after detrending range from 24 to 38 ppm, and are generally slightly smaller when T$_0$ is fitted.
Apart from stare-mode visit 2, the fixed and fitted T$_0$ agree within 2 minutes. and the corresponding transit radii Rp/Rs agree within the uncertainties.

\begin{figure}[t!]
\centering
\includegraphics[scale=0.15]{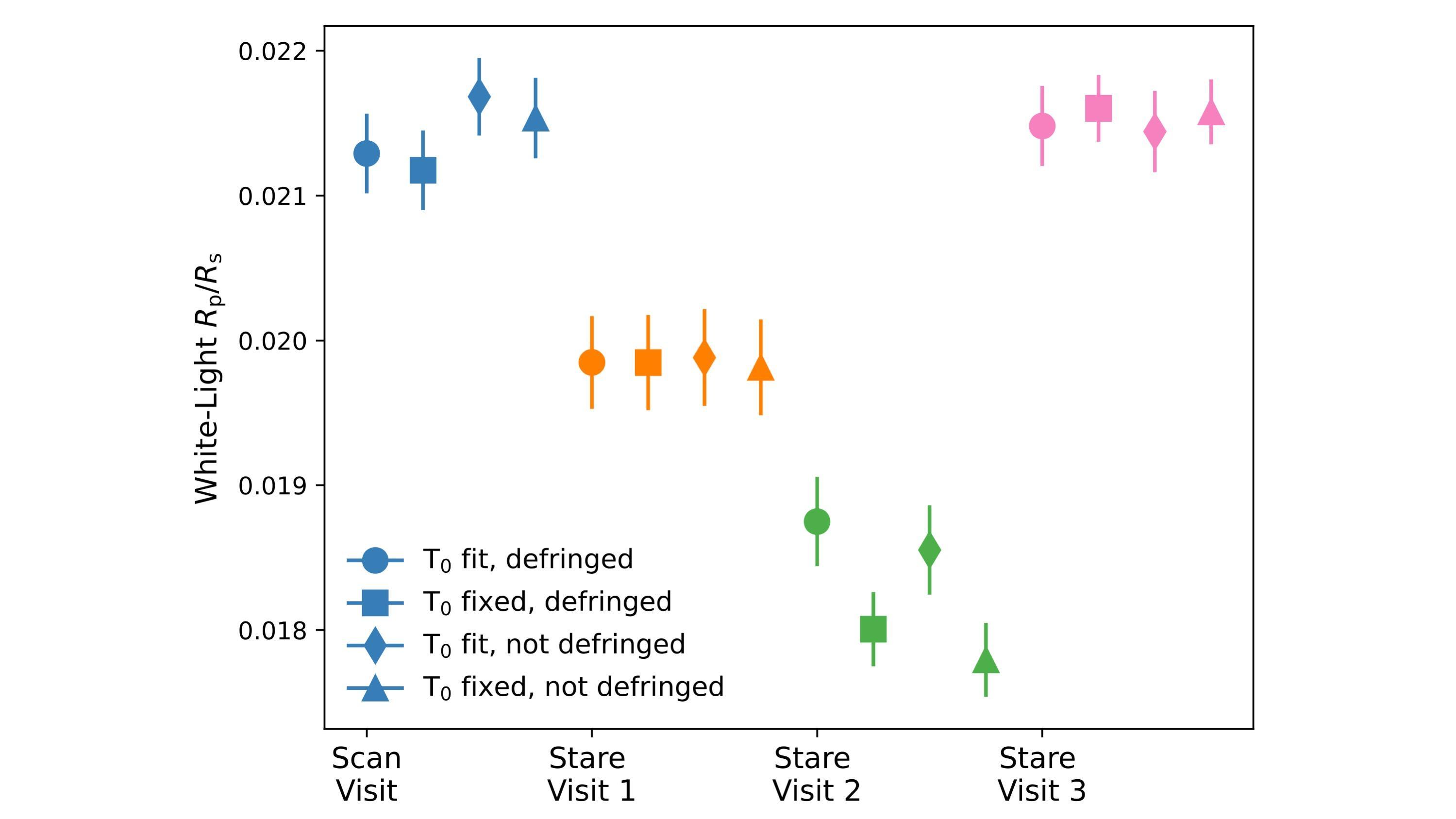}
\caption{{\small Comparison of measured white-light curve transit radii for each of the visits from programs 16442 (scan-mode) and 13665 (stare-mode), for the four reduction scenarios.}}
\label{fig:jl8}
\end{figure}

The transit radii [Rp/Rs = (transit depth)$^{1/2}$] derived from the white-light curves are shown in Figure~\ref{fig:jl8}.
In general, the transit depths in the defringed white-light curves are consistent with the transit depths in the non-defringed white-light curves.
The transit depths between fixed and fitted transit center also seem to be consistent, except for visit 2 of the stare-mode data, where the fitted T$_0$ is $\sim$26 minutes less than the ephemeris value.
(That difference may be due to the slightly higher flux for the first few exposures in the second (third) orbit of that visit, which the variable-T$_0$ detrending fit appears to have associated with the end of ingress.)
The most striking feature in comparing the various curves, however, is the disagreement in the transit depths among the individual visits -- by more than 3$\sigma$ for the three stare-mode visits.
The minimum radius measured (from stare-mode visit 2) corresponds to a radius of approximately 1.8 Earth-radii, compared to a radius of 2.2 Earth-radii in the scan-mode visit and stare-mode visit 3.

\begin{figure}[t!]
\centering
\includegraphics[scale=0.16, clip = true, trim = 0 225 0 0]{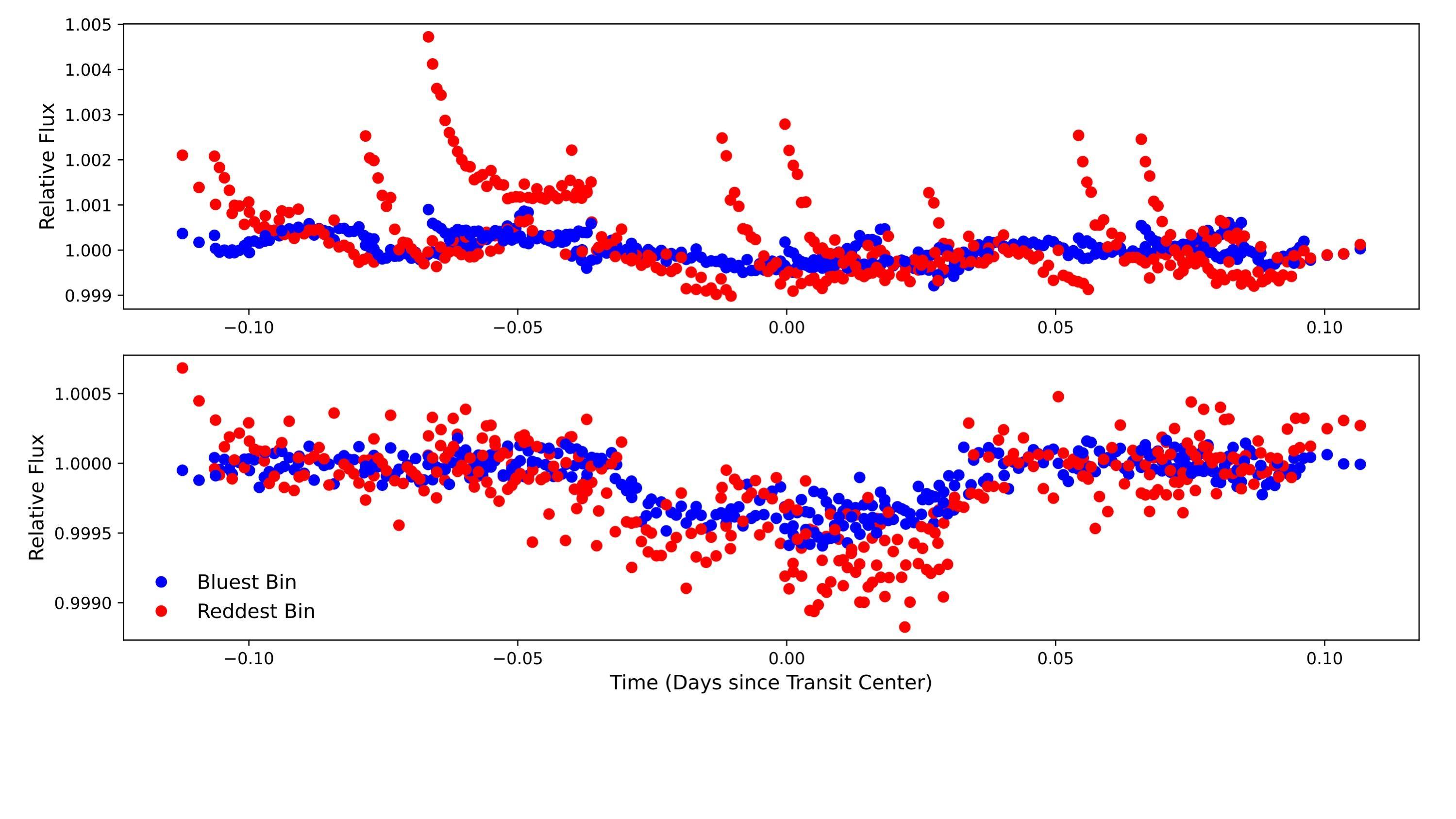}
\caption{{\small Comparison of the raw ({\it upper panel}) and detrended ({\it lower panel}) light curves for the bluest and reddest spectroscopic bins (i.e., 5500-6000 \AA\ and 9500-10000 \AA, respectively).
Data from all visits from programs 16442 (scan-mode) and 13665 (stare-mode) are included, with T$_0$ fixed to the ephemeris.
The transit is deeper for the reddest spectral bin.}}
\label{fig:jl11}
\end{figure}
\begin{figure}[b!]
\centering
\includegraphics[scale=0.16, clip = true, trim = 0 300 0 275]{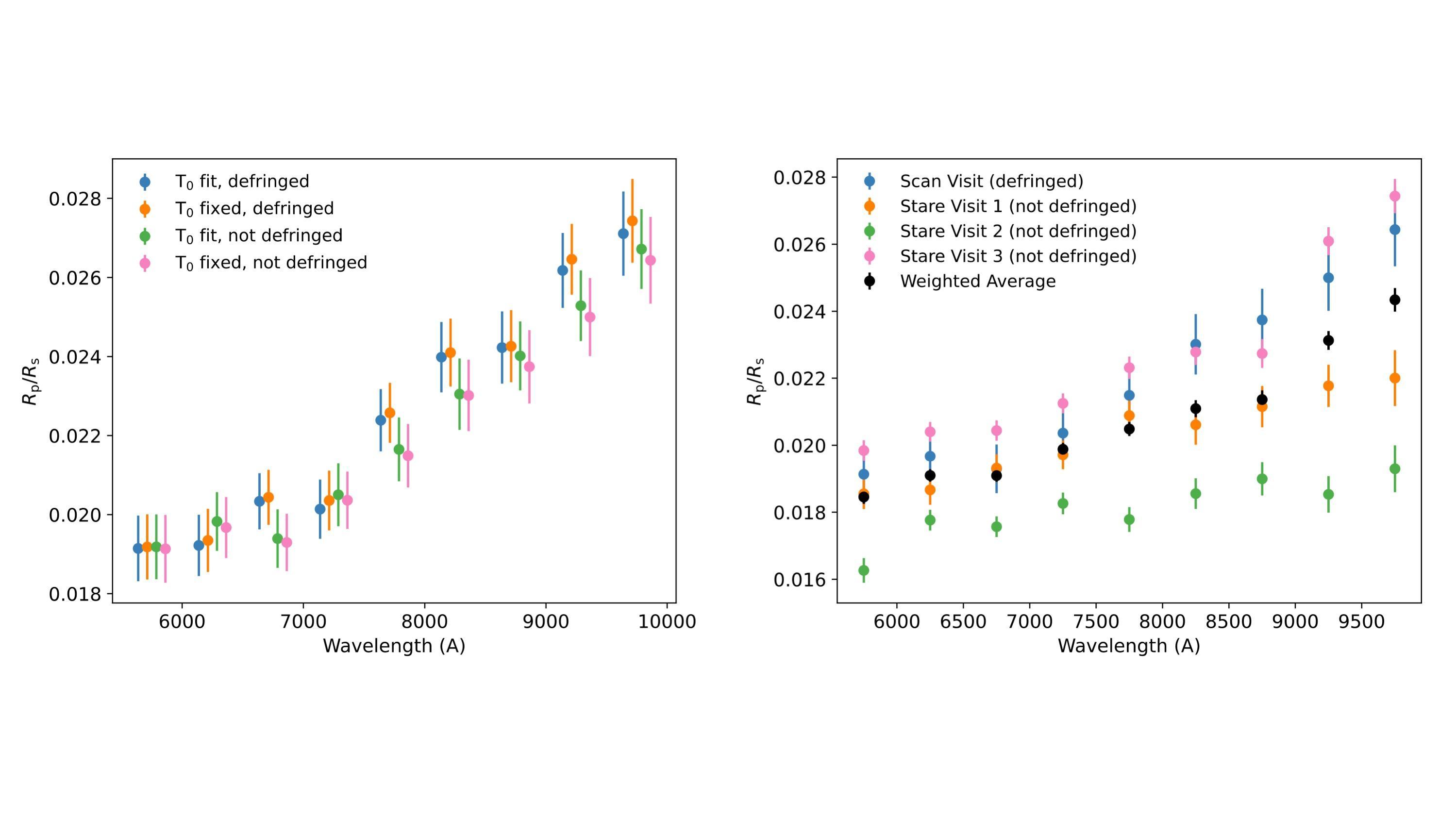}
\caption{{\small ({\it left}) Transit radii [Rp/Rs = (transit depth)$^{1/2}$] from the scan-mode spectra (program 16442), for different reduction scenarios. 
Data points are offset in wavelength for clarity.
({\it right}) Transit radii for scan-mode spectra (program 16442; in blue) and stare-mode spectra (program 13665).
Only the scan-mode spectra were defringed. 
Here the transit center is fixed to the ephemeris of \href{https://ui.adsabs.harvard.edu/abs/2023ApJS..265....4K/abstract}{Kokori et al. (2023)}.
For all four data sets, the transit depth appears to increase with wavelength -- though the slope of that increase differs among the four.}}
\label{fig:jl910}
\end{figure}

After considering the white-light curves, we also split the spectra into nine spectroscopic bins, evenly spaced between 5500 and 10000 \AA, and fit each light curve with the full systematics model, with no common mode correction (\href{https://ui.adsabs.harvard.edu/abs/2012MNRAS.422.2477H/abstract}{Huitson et al. 2012}).
Some details regarding the coefficients used in the model fits are given in Appendix A.
Figure~\ref{fig:jl11} shows the raw and detrended light curves for the bluest and reddest spectroscopic bins.
In the detrended light curve the transit depths are clearly deeper at the redder wavelengths, despite the larger statistical noise from the decreased throughput there.
The left-hand panel of Figure~\ref{fig:jl910} shows the transit radii derived from the scan-mode data, for the different analysis scenarios from Figure~\ref{fig:jl8} (i.e., fixed versus fitted T$_0$ and defringing the data (or not)).
In all cases, a significant redward slope can be clearly seen, with the transit radius for the reddest bin (at about 1 $\mu$m) more than 40\% larger than the transit radius for the bluest bin (around 0.6 $\mu$m).
Notably, this slope towards larger transit depths at longer wavelengths is seen in each of the stare-mode visits as well, as shown in the right-hand panel of Figure~\ref{fig:jl910}, suggesting that this behavior is not due to the spatial scan observational strategy, nor to the strong saturation in the stare-mode data.
While each visit exhibits a clear longward slope, the magnitude of the slope appears to vary from visit to visit, with the scan-mode visit agreeing most closely with visit 3 of the stare-mode observations (as was the case for the white-light transit depths).
Visit 1 of the stare-mode data exhibits a shallower slope, with visit 2 shallower still.

55 Cancri e has long been studied for possible time variability in its transit depths, occultation depths, and phase curve (e.g., \href{https://ui.adsabs.harvard.edu/abs/2018AJ....155..221T/abstract}{Tamburo et al. 2018}; \href{https://ui.adsabs.harvard.edu/abs/2019A&A...631..129S/abstract}{Sulis et al. 2019}).
\href{https://ui.adsabs.harvard.edu/abs/2016MNRAS.455.2018D/abstract}{Demory et al. (2016)} found marginal variability in the depths among six transits observed with Spitzer.
\href{https://ui.adsabs.harvard.edu/abs/2023A&A...677..112M/abstract}{Meier Valdes et al. (2023)}, however, found consistent transit depths (within uncertainties) for phase-folded TESS transits from sectors 21, 44, and 46, and \href{https://ui.adsabs.harvard.edu/abs/2024AJ....167....1W/abstract}{Wang \& Espinoza (2024)} found no evidence of variability over the course of 88 individual TESS transit observations.
Those prior results suggest that the white-light transit depth differences in our four visits might not reflect true astrophysical variability, but might instead be due either to the phasing of the observations (with incomplete and non-uniform sampling of the transits due to the gaps from Earth occultation) or to some as yet unrecognized instrumental effect.
On the other hand, recent JWST/NIRCam observations of five occultations of 55 Cnc e -- while not directly comparable to the transit observations discussed here -- have revealed apparent variability in the optical/infrared occultation depths, which might be due to variability in a thin CO/CO$_2$ atmosphere (\href{https://ui.adsabs.harvard.edu/abs/2024A&A...690A.159P/abstract}{Patel et al. 2024}; \href{https://ui.adsabs.harvard.edu/abs/2026arXiv260611866S/abstract}{Snellen et al. 2026}).

Even if the mean transit depths are not as variable as our analysis suggests, however, the positive slope toward longer wavelengths appears to be robust among all visits and reductions.
Interestingly, the systematic trends in the reddest bin appear much larger than those at bluer wavelengths, though it is unclear how those larger systematic trends might translate to larger transit depths, or what their physical cause might be.
In any case, the results for the scan-mode visit suggest that the positive slope towards longer wavelengths is not due to the saturation of the STIS CCD, as the scanned observations were well under the saturation limit.
The longward slope also does not appear to be due to fringing at the longer wavelengths covered by the G750L setting, as the slope appears to be consistent among the different reductions (Figure~\ref{fig:jl910}).


\ssection{Summary}\label{sec-sum}

We have discussed STIS G750L spectra of the known exoplanet host 55 Cnc, obtained under two special calibration programs that were designed to assess the utility of STIS spatial scans for time-series observations of transiting exoplanets -- and for other applications requiring very high-S/N optical/near-IR spectra and/or very accurate, reproducible fluxes.
The first visit of program 15383 performed some initial tests of the scan geometry and observational strategies -- leading to recommendations of a slight adjustment of the scan angle (to optimize the defringing) and to only use forward scans.
The second visit obtained a series of 20 sub-array scans of 55 Cnc, with the super-Earth 55 Cnc e out of transit, to test the reproducibility of the fluxes in that time series and to experiment with methods for cosmic-ray correction, defringing, extraction of 1D spectra, and removal of additional subtle observational/instrumental effects (detrending).
Program 16442 obtained a series of 42 very similar sub-array scans of 55 Cnc, covering a transit of 55 Cnc e, in order to refine the analysis methods and to see how well the transit could be characterized.
Deliberately saturated (``stare-mode'') STIS spectra of 55 Cnc from the MAST archive (program 13665), which covered three earlier transits of 55 Cnc e using the same instrumental setup, were also analyzed for comparison. 
We adopted the L.A.C{\sc osmic} package for cosmic-ray correction, the new Python version of the {\bf defringe} tool in {\bf stistools}, and a detrending approach commonly used for the analysis of exoplanet transit data.

The analysis of the two epochs of scan-mode spectra, and comparisons with the three additional epochs of stare-mode spectra have yielded both useful guidance for future observations and some characteristics of the transit of 55 Cnc e:
\begin{itemize}
\item{The systematic differences in flux between orbits (before detrending) -- of order several hundred ppm for the scan-mode data -- are smaller than those typically seen for pointed/saturated exposures (which can be as high as several thousand ppm).
Within each orbit, the differences in flux among the scanned spectra are also generally smaller than those often seen for the stare-mode exposures -- though part of that may be due to the scan-mode observations averaging over longer exposure times and larger regions on the CCD detector.}
\item{The trends in the relative fluxes can differ somewhat with wavelength.}
\item{For the total (``white-light'') flux, the scatter about the detrending fits can be as low as $\sim$30 ppm – comparable to the best values achieved in previous studies of exoplanet host stars with HST.
Somewhat larger values are found for narrower wavelength regions (particularly at the longer wavelengths).}
\item{While averaging over the larger spatial range in the scanned spectral images does reduce the amplitude of the fringing in 1D spectra extracted from those images, the additional step of defringing the images both effectively removes the fringes and appears to reduce the scatter about the detrending fits by of order 15-20\% at the longer wavelengths.
The higher rms values there reflect the lower count rates at the redder wavengths.}
\item{The white-light transit depth for 55 Cnc e derived from the scan-mode data is of order 450 ppm -- comparable to previous values determined from optical spectra.
The three stare-mode observations of this system yield somewhat different values for the transit depth, however.}
\item{The transit radii (Rp/Rs) appear to exhibit an unexpected increase for wavelength intervals from 5500 \AA\ to 1 $\mu$m -- both for the scanned spectrum and for the three stare-mode spectra.
The roughly 40\% increase in Rp/Rs for the scan-mode spectrum corresponds to a facor of $\sim$2 increase in the transit depth.
The slopes of the increase differ among the four epochs probed, however -- particularly among the three stare-mode spectra -- which may suggest an instrumental origin for the increase.
Upcoming observations with JWST NIRISS/SOSS (program GO-12237; V. Boehm, PI) should provide more definitive information regarding the possible temporal and wavelength dependence of the transit depth in this system.}
\end{itemize}

While these results are based on somewhat limited data (20+42 scanned spectra obtained during just two+five orbits in two HST visits), they do indicate that spatial scanning with the STIS CCD might be advantageous for some applications, compared to other available HST observing modes.
For example, while WFC3, with the UVIS and IR grisms, can provide high S/N spectra of fainter targets with broader wavelength coverage, STIS spatial scans may be preferable for obtaining higher resolution spectra of brighter targets (e.g., for V $<$ 7.5) – particularly at wavelengths greater than about 6000 \AA, where the UVIS spectra will exhibit saturation and contamination from overlapping orders and the IR spectra would require rather rapid scan rates.

Observers interested in using spatial scans may consult \href{https://hst-docs.stsci.edu/stisihb/chapter-12-special-uses-of-stis/12-12-spatial-scans-with-the-stis-ccd}{Section 12.12 in the STIS Instrument Handbook} for instructions on how to construct the phase II proposals and how to estimate exposure times for both science targets and fringe-flat exposures.
For any given target, adjustment of the scan rate and duration will need to be carefully assessed, in order to achieve (if possible) both high enough S/N (without saturating the CCD) and adequate temporal sampling to fulfill the science goals.
Additional information may be found in the phase II proposals for the programs that have used spatial scans so far: M. Cordiner's programs using G750M (14705, 15429, 15478); the special calibration programs using G750L discussed in this report (15383, 16442); and a recent exoplanet study using scans with G430L (17537; G. Fu, PI).


\ssectionstar{Acknowledgements}\label{sec:Ackn}
We thank John Debes for bringing us together to work on this project and for reviewing this report, Sean Lockwood for obtaining the high-resolution jitter files, and Susana Deustua for early work on the cosmic ray correction.
Those three, and N\'{e}stor Espinoza, also contributed many valuable suggestions during the early stages of the project.

This work is based on observations with the NASA/ESA Hubble Space Telescope, obtained at the STScI operated by AURA, Inc. -- including archival data from program 13365 (B. Benneke, PI).

This work made use of the following software packages:  {\sc iraf}, IDL Astronomy user’s library (\href{https://ui.adsabs.harvard.edu/abs/1995ASPC...77..437L/abstract}{Landsman 1995}), NumPy (Oliphant 2006), SciPy (Virtanen et al. 2019), MatPlotLib (Caswell et al. 2019), {\bf stistools}, L.A.C{\sc osmic} (\href{https://ui.adsabs.harvard.edu/abs/2001PASP..113.1420V/abstract}{van Dokkum 2001}), Exoplanet Characterization Toolkit (\href{https://ui.adsabs.harvard.edu/abs/2021zndo...4556063B/abstract}{Bourque et al. 2021})

\vspace{-0.3cm}
\ssectionstar{Change History for STIS ISR 2026-05}\label{sec:History}
\vspace{-0.3cm}
Version 1: \ddmonthyyyy\date{21 August 2026} - Original Document 

\vspace{-0.3cm}
\ssectionstar{References}\label{sec:References}
\vspace{-0.3cm}

\noindent
\href{https://ui.adsabs.harvard.edu/abs/2010PASP..122.1035A/abstract}{Anderson, J., \& Bedin, L. R. 2010}, PASP, 122, 1035    

\noindent
\href{https://www.stsci.edu/files/live/sites/www/files/home/hst/instrumentation/stis/documentation/instrument-science-reports/_documents/1997_15.pdf}{Baum, S., Ferguson, H., Walsh, J. R.,  et al. 1998}, STIS ISR 1997-15, GO Added Near-IR Fringe Flats (Rev. A)

\noindent
\href{https://ui.adsabs.harvard.edu/abs/2004AJ....127.3508B/abstract}{Bohlin, R. C., \& Gilliland, R. L. 2004}, AJ, 127, 3508

\noindent
\href{https://ui.adsabs.harvard.edu/abs/2021zndo...4556063B/abstract}{Bourque, M., Espinoza, N., Filippazzo, J., et al. 2021}, https://doi.org/10.5281/zenodo.4556063  

\noindent
\href{https://ui.adsabs.harvard.edu/abs/2018A&A...619....1B/abstract}{Bourrier, V., Dumusque, X., Dorn, C., et al. 2018}, A\&A, 619A, 1    

\noindent
\href{https://ui.adsabs.harvard.edu/abs/2001ApJ...552..699B/abstract}{Brown, T. M., Charbonneau, D., Gilliland, R. L., et al. 2001}, ApJ, 552, 699    

\noindent
Caswell, T. A., Droettboom, M., Hunter, J., et al. 2019, matplotlib/matplotlib v3.1.0, v3.1.0, Zenodo (doi: 10.5281/zenodo.2893252)

\noindent
\href{https://ui.adsabs.harvard.edu/abs/2017ApJ...843L...2C/abstract}{Cordiner, M. A., Cox, N. L. J., Lallement, R., et al. 2017}, ApJL, 843, L2

\noindent
\href{https://ui.adsabs.harvard.edu/abs/2019ApJ...875L..28C/abstract}{Cordiner, M. A., Linnartz, H., Cox, N. L. J., et al. 2019}, ApJL, 875, L28

\noindent
\href{https://ui.adsabs.harvard.edu/abs/2010ApJ...722..937D/abstract}{Dawson, R. I. \& Fabrycky, D. C. 2010}, ApJ, 722, 937     

\noindent
\href{https://ui.adsabs.harvard.edu/abs/2013ApJ...774...95D/abstract}{Deming, D., Wilkins, A., McCullough, P., et al. 2013}, ApJ, 774, 95     

\noindent
\href{https://ui.adsabs.harvard.edu/abs/2015MNRAS.450.2043D/abstract}{Demory, B.-O., Ehrenreich, D., Queloz, D., et al. 2015}, MNRAS, 450, 2043     

\noindent
\href{https://ui.adsabs.harvard.edu/abs/2016MNRAS.455.2018D/abstract}{Demory, B.-O., Gillon, M., Madhusudhan, N., et al. 2016}, MNRAS, 455, 2018     

\noindent
\href{https://ui.adsabs.harvard.edu/abs/2013ApJ...772L..16E/abstract}{Evans, T. M., Pont, F., Sing, D. K., et al. 2013}, ApJL, 772, L16

\noindent
\href{https://ui.adsabs.harvard.edu/abs/2008ApJ...675..790F/abstract}{Fischer, D. A., Marcy, G. W., Butler, R. P., et al. 2008}, ApJ, 675, 790

\noindent
\href{https://ui.adsabs.harvard.edu/abs/1999PASP..111.1009G/abstract}{Gilliland, R. L., Goudfrooij, P., \& Kimble, R. A. 1999}, PASP, 111, 1009

\noindent
\href{https://www.stsci.edu/files/live/sites/www/files/home/hst/instrumentation/stis/documentation/instrument-science-reports/_documents/2006_03.pdf}{Goudfrooij, P., \& Bohlin, R. C. 2006}, STIS ISR 2006-03, A new CTE Correction Algorithm for Point Source Spectroscopy with the STIS CCD

\noindent
\href{https://ui.adsabs.harvard.edu/abs/2006PASP..118.1455G/abstract}{Goudfrooij, P., Bohlin, R. C., Ma\'{i}z-Apell\'{a}niz, J., \& Kimble, R. A. 2006}, PASP, 118, 1455    

\noindent
\href{https://www.stsci.edu/files/live/sites/www/files/home/hst/instrumentation/stis/documentation/instrument-science-reports/_documents/1998_19.pdf}{Goudfrooij, P., Bohlin, R. C., Walsh, J. R., \& Baum, S. A. 1998}, STIS ISR 1998-19, STIS Near-IR Fringing. II. Basics and Use of Contemporaneous Flats for Spectroscopy of Point Sources

\noindent
\href{https://www.stsci.edu/files/live/sites/www/files/home/hst/instrumentation/stis/documentation/instrument-science-reports/_documents/1998_29.pdf}{Goudfrooij, P., \& Christensen, J. A. 1998}, STIS ISR 1998-29, STIS Near-IR Fringing. III. A tutorial on the Use of the {\sc iraf} Tasks

\noindent
\href{https://ui.adsabs.harvard.edu/abs/2020AJ....159..239G/abstract}{Guo, X., Crossfield, I. J. M., Dragomir, D., et al. 2020}, AJ, 159, 239

\noindent
\href{https://ui.adsabs.harvard.edu/abs/2012MNRAS.422.2477H/abstract}{Huitson, C. M., Sing, D. K., Vidal-Madjar, A. et al. 2012}, MNRAS, 422, 2477

\noindent
\href{https://ui.adsabs.harvard.edu/abs/2013A&A...553....6H/abstract}{Husser, T.-O., Wende-von Berg, S., Dreizler, S.,  et al. 2013}, A\&A, 553A, 6  

\noindent
\href{https://ui.adsabs.harvard.edu/abs/2023ApJS..265....4K/abstract}{Kokori, A., Tsiaras, A., Edwards, B.,  et al. 2023}, ApJS, 265, 4  

\noindent
\href{https://ui.adsabs.harvard.edu/abs/2015PASP..127.1161K/abstract}{Kreidberg, L. 2015}, PASP, 127, 1161  

\noindent
\href{https://ui.adsabs.harvard.edu/abs/1995ASPC...77..437L/abstract}{Landsman, W. B. 1995}, Astronomical Society of the Pacific Conference Series, Vol. 77, The IDL Astronomy User's Library, ed. R. A. Shaw, H. E. Payne, \& J. J. E. Hayes, 437

\noindent
\href{https://ui.adsabs.harvard.edu/abs/2008A&A...481...83L/abstract}{Lecavelier des Etangs, A., Pont, F., Vidal-Madjar, A., et al. 2008}, A\&A, 481, L83  

\noindent
\href{https://ui.adsabs.harvard.edu/abs/2018AJ....155...66L/abstract}{Lothringer, J. D., Benneke, B., Crossfield, I. J. M., et al. 2018}, AJ, 155, 66    

\noindent
\href{https://www.stsci.edu/files/live/sites/www/files/home/hst/instrumentation/stis/documentation/instrument-science-reports/_documents/2019_01.pdf}{Maclay, M. T., \& Debes, J, H. 2019}, STIS ISR 2019-01, A New Method to Monitor the HST/STIS Focus

\noindent
\href{https://www.stsci.edu/files/live/sites/www/files/home/hst/instrumentation/wfc3/documentation/instrument-science-reports/_documents/2012_08.pdf}{McCullough, P., \& MacKenty, J. 2012}, WFC3 ISR 2012-08, Considerations for using Spatial Scans with WFC3

\noindent
\href{https://ui.adsabs.harvard.edu/abs/2014ApJ...791...55M/abstract}{McCullough, P. R., Crouzet, N., Deming, D., et al. 2014}, ApJ, 791, 55  

\noindent
\href{https://ui.adsabs.harvard.edu/abs/2023A&A...677..112M/abstract}{Meier Valdes, E. A., Morris, B. M., Demory, B.-O., et al. 2023}, A\&A, 677A, 112

\noindent
\href{https://ui.adsabs.harvard.edu/abs/2014MNRAS.437...46N/abstract}{Nikolov, N., Sing, D. K., Pont, F., et al. 2014}, MNRAS, 437, 46    

\noindent
Oliphant, T. 2006, NumPy: A guide to NumPy, USA: Trelgol Publishing (http://www.numpy.org)

\noindent
\href{https://ui.adsabs.harvard.edu/abs/2014A&A...568...99O/abstract}{Oshagh, M., Santos, N. C., Ehrenreich, D., et al. 2014}, A\&Ap, 568A, 99  

\noindent
\href{https://ui.adsabs.harvard.edu/abs/2024A&A...690A.159P/abstract}{Patel, J. A., Brandeker, A., Kitzmann, D., et al. 2024}, A\&Ap, 690A, 159  

\noindent
\href{https://www.stsci.edu/files/live/sites/www/files/home/hst/instrumentation/stis/documentation/instrument-science-reports/_documents/2015_06.pdf}{Proffitt, C. R. 2015}, STIS ISR 2015-06, CCD saturation effects

\noindent
\href{https://www.stsci.edu/files/live/sites/www/files/home/hst/instrumentation/stis/documentation/instrument-science-reports/_documents/2017_01.pdf}{Proffitt, C. R., Monroe, T., \& Dressel, L. 2017}, STIS ISR 2017-01, Status of the STIS Instrument Focus

\noindent
\href{https://www.stsci.edu/files/live/sites/www/files/home/hst/instrumentation/stis/documentation/instrument-science-reports/_documents/2018_06.pdf}{Riley, A., Monroe, T., \& Lockwood, S. 2018}, STIS ISR 2018-06, Impacts of focus on aspects of STIS UV spectroscopy

\noindent
\href{https://www.stsci.edu/files/live/sites/www/files/home/hst/instrumentation/wfc3/documentation/instrument-science-reports/_documents/2017_21.pdf}{Shanahan, C. E., McCullough, P., \& Baggett, S. 2017}, WFC3 ISR 2017-21, Photometric Repeatability of Scanned Imagery:  UVIS

\noindent
\href{https://www.stsci.edu/files/live/sites/www/files/home/hst/instrumentation/stis/documentation/instrument-science-reports/_documents/1998_22.pdf}{Shaw, R., \& Hodge, P. 1998}, STIS ISR 1998-22, Cosmic Ray Rejection in STIS CCD Images

\noindent
\href{https://www.stsci.edu/files/live/sites/www/files/home/hst/instrumentation/stis/documentation/instrument-science-reports/_documents/1998_11.pdf}{Shaw, R., \& Hsu, J. C. 1998}, STIS ISR 1998-11, Calstis2:  Cosmic Ray Rejection in the STIS Calibration Pipeline

\noindent
\href{https://arxiv.org/abs/1804.07357}{Sing, D. K. 2018}, arXiv e-prints, arXiv:1804.07357

\noindent
\href{https://ui.adsabs.harvard.edu/abs/2011A&A...527...73S/abstract}{Sing, D. K., D\'{e}sert, J.-M., Fortney, J. J., et al. 2011}, A\&A, 527, A73 (doi: 10.1051/0004-6361/201015579)

\noindent
\href{https://ui.adsabs.harvard.edu/abs/2019AJ....158...91S/abstract}{Sing, D. K., Lavras, P., Ballester, G. E., et al. 2019}, AJ, 158, 91 (doi: 10.3847/1538-3881/ab2986)   

\noindent
\href{https://ui.adsabs.harvard.edu/abs/2013MNRAS.436.2956S/abstract}{Sing, D. K., Lecavelier des Etangs, A., Fortney, J. J., et al. 2013}, MNRAS, 436, 2956    

\noindent
\href{https://ui.adsabs.harvard.edu/abs/2008ApJ...686..658S/abstract}{Sing, D. K., Vidal-Madjar, A., D\'{e}sert, J.-M., et al. 2008}, ApJ, 686, 658

\noindent
\href{https://ui.adsabs.harvard.edu/abs/2026arXiv260611866S/abstract}{Snellen, I., Miguel, Y., Janssen, L., et al. 2026}, arXiv:2606.11866   

\noindent
\href{https://ui.adsabs.harvard.edu/abs/2019A&A...631..129S/abstract}{Sulis, S., Dragomir, D., Lendl, M., et al. 2019}, A\&A, 631, A129   

\noindent
\href{https://ui.adsabs.harvard.edu/abs/2018AJ....155..221T/abstract}{Tamburo, P., Mandell, A., Deming, D., \& Garhart, E. 2018}, AJ, 155, 221

\noindent
\href{https://ui.adsabs.harvard.edu/abs/2016ApJ...832..202T/abstract}{Tsiaras, A., Waldmann, I. P., Rocchetto, M., et al. 2016}, ApJ, 832, 202   

\noindent
\href{https://ui.adsabs.harvard.edu/abs/2001PASP..113.1420V/abstract}{van Dokkum, P. G. 2001}, PASP, 113, 1420   

\noindent
Virtanen, P., Gommers, R., Burovski, E., et al. 2019, scipy/scipy: SciPy 1.2.1, v1.2.1, Zenodo (doi: 10.5281/zenodo.2560881)

\noindent
\href{https://ui.adsabs.harvard.edu/abs/2016ApJ...819...10W/abstract}{Wakeford, H. R., Sing, D. K., Evans, T., et al. 2016}, ApJ, 819, 10    

\noindent
\href{https://www.stsci.edu/files/live/sites/www/files/home/hst/instrumentation/stis/documentation/instrument-science-reports/_documents/1997_16.pdf}{Walsh, J. R., Baum, S. A., Malamuth, E. M., \& Goudfrooij, P. 1997}, STIS ISR 1997-16, STIS Near-IR Fringing:  Basics and Use of Contemporaneous Flats for Extended Sources

\noindent
\href{https://ui.adsabs.harvard.edu/abs/2024AJ....167....1W/abstract}{Wang, G., \& Espinoza, N. 2024}, AJ, 167, 1

\noindent
\href{https://ui.adsabs.harvard.edu/abs/2011ApJ...737L..18W/abstract}{Winn, J. N., Matthews, J. M., Dawson, R. I., et al. 2011}, ApJL, 737, L18   


\clearpage
\ssectionstar{Appendix A -- Image parameters and detrending coefficients}\label{sec:AppA}
\vspace{-0.3cm}
Table~\ref{tab:parms} lists several parameters related to the position of the 2D images on the detector, for the 20 scanned spectra obtained in visit 2 of program 15383.
The zero point of the HST orbital phase is set by the first exposure.
The focus values, taken from the UVIS1 models, indicate a systematic decline through each visibility period (toward better focus for STIS).
Several measures of the relative x-position are given:  from cross correlation of rows (spectrum vs. fringe flat), from the defringing procedure, and from independent cross correlations used for input to the detrending.
All three of those measures (and the focus) appear to be well correlated, modulo zero points and scale factors.
The factors used to scale the fringe flats in the defringing are all $\sim$0.9. 
 
\begin{figure}[b!]
\centering
\begin{minipage}[c]{0.3\textwidth}
   \includegraphics[scale=0.5]{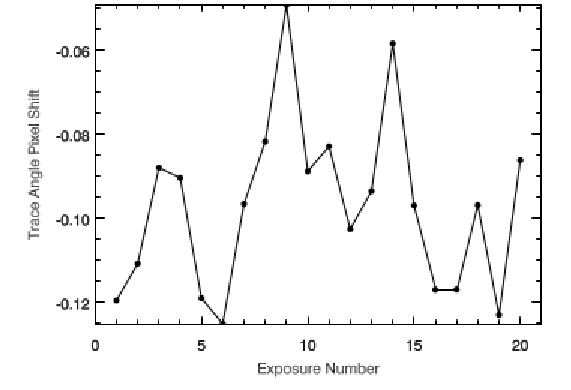}
\end{minipage} \hfill
\begin{minipage}[c]{0.3\textwidth}
   \includegraphics[scale=0.5]{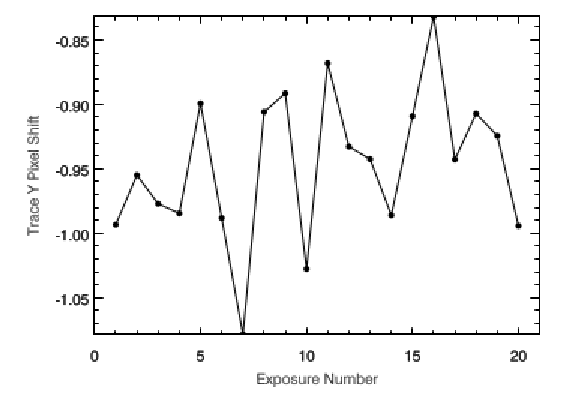}
\end{minipage} \hfill
\begin{minipage}[c]{0.3\textwidth}
   \includegraphics[scale=0.5]{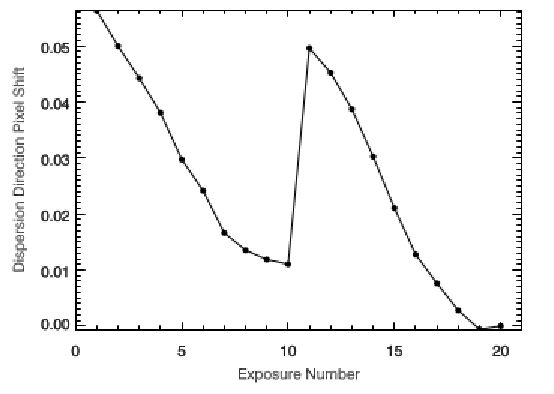}
\end{minipage}
\caption{{\small Several of the parameters used in exploring the detrending of the 20 scans from visit 2 of program 15383: ({\it left}) relative angle of the central aperture trace, ({\it center}) relative y pixel of the central aperture trace, ({\it right}) wavelength shift measured from cross correlation.
Note:  these are for the flt files, which are not corrected for cosmic rays or fringing.}}
\label{fig:detrparms}
\end{figure}

The parameters adopted for the exploratory systematics model used for detrending the scan-mode data obtained in visit 2 of program 15383 included the time, the HST orbital phase (to 4th order), the relative angle of the central aperture trace, the relative pixel of the central aperture trace (in y), and the wavelength shift.
The latter three, which were determined via explicit tracing of the scanned spectra in the 2D flt images and subsequent cross-correlations among the 1D extracted spectra, are shown in Figure~\ref{fig:detrparms}.
There is a clear trend in the wavelength shifts -- declining by $\sim$0.05 pix through each visibility period, and slightly offset between the two orbits, and perhaps a slight increase in the y-location of the spectral trace over the two orbits (by $\sim$0.05 pix).

\begin{figure}
\centering
\includegraphics[scale=0.28]{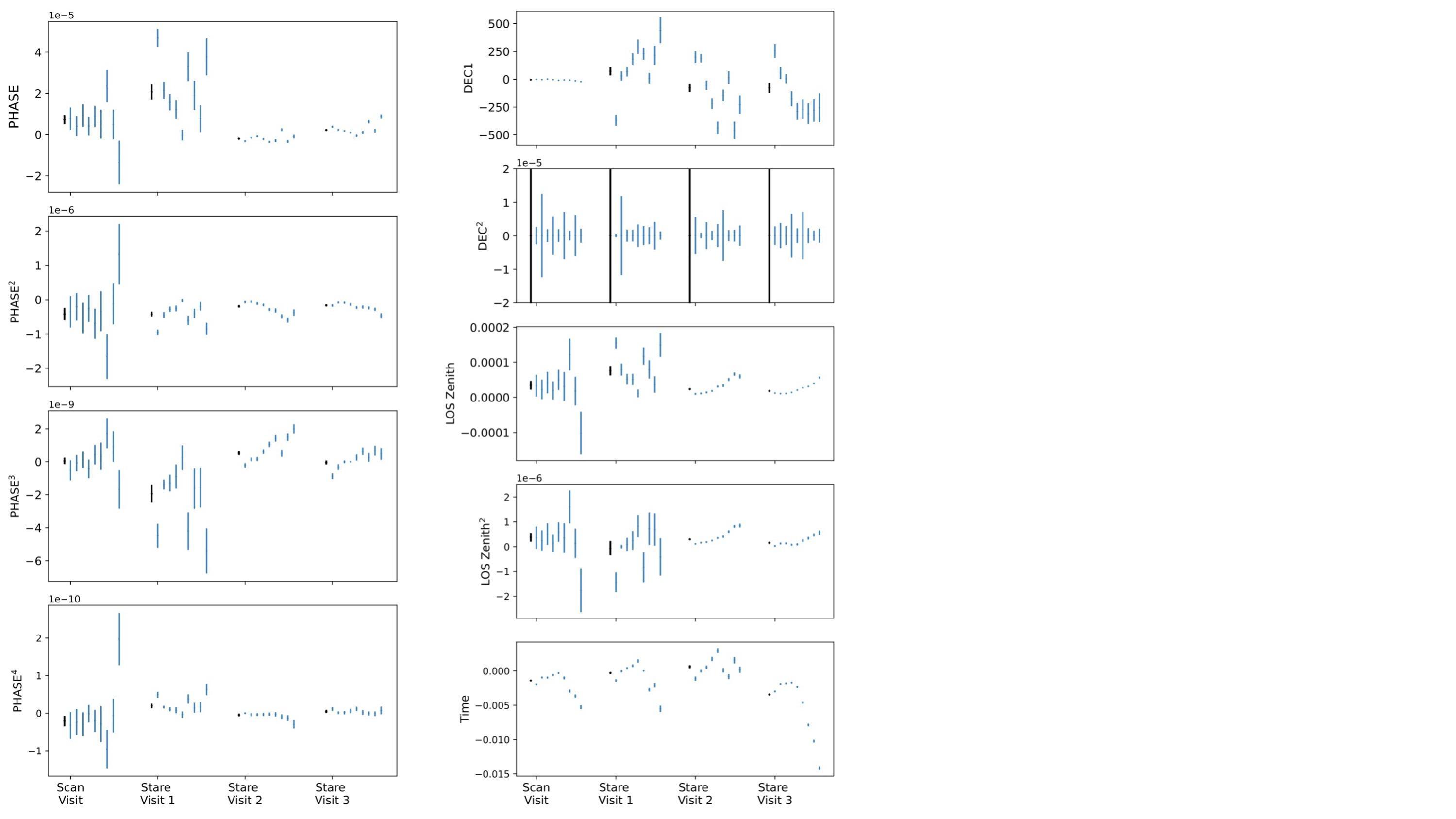}
\caption{{\small Coefficients of the parameters used in the systematics model fits to the white-light (black) and spectroscopic (blue) light curves, for the scan-mode (program 16442) and stare-mode (program 13665) data.}}
\label{fig:jl45}
\end{figure}

The parameters adopted for the model used for detrending the scan-mode and stare-mode data covering a transit of 55 Cnc e included the time, the HST orbital phase (to 4th order), the declination (to 2nd order), and the line of sight zenith angle (to 2nd order).
Figure~\ref{fig:jl45} shows the coefficients of those terms, for the final model fits to both the white-light and spectroscopic light curves, for all four of the visits covering a transit of 55 Cnc e.
In each panel, the coefficients for the white-light fluxes are given in black, and the coefficients for the fluxes in the nine narrower spectral intervals between 0.55 and 1.0 $\mu$m are given in blue.
For most of the parameters, the uncertainties in the coefficients are somewhat larger for the scan-mode data, and smallest for stare-mode visits 2 and 3.
The striking similarity between the coefficients for stare-mode visits 2 and 3 may be a consequence of those observations having been obtained fairly close in time (within 2 weeks of each other).
The larger number of exposures obtained during stare-mode visits 2 and 3 (which used sub-arrays to increase observing efficiency) may contribute to the generally smaller uncertainties found for those two visits.

\begin{deluxetable}{rrrrrrr}
\tablecolumns{7}
\tabletypesize{\footnotesize}
\tablecaption{Image parameters for visit 2 of program 15383\label{tab:parms}}
\tablewidth{0pt}

\tablehead{
\multicolumn{1}{c}{Exposure}&
\multicolumn{1}{c}{HST phase\tablenotemark{a}}&
\multicolumn{1}{c}{focus\tablenotemark{b}}&
\multicolumn{1}{c}{xcorr\tablenotemark{c}}&
\multicolumn{2}{c}{Defringing\tablenotemark{d}}&
\multicolumn{1}{c}{Detrending\tablenotemark{e}}\\
\multicolumn{4}{c}{ }&
\multicolumn{1}{c}{xoff}&
\multicolumn{1}{c}{scale}&
\multicolumn{1}{c}{disp off}\\
\multicolumn{1}{c}{(1)}&
\multicolumn{1}{c}{(2)}&
\multicolumn{1}{c}{(3)}&
\multicolumn{1}{c}{(4)}&
\multicolumn{1}{c}{(5)}&
\multicolumn{1}{c}{(6)}&
\multicolumn{1}{c}{(7)}}

\startdata
 1 &   0.0000 &   2.37 & $-$0.121 & $-$0.196 & 0.880 &   1.295 \\
 2 &   0.0507 &   1.57 & $-$0.180 & $-$0.302 & 0.879 &   1.001 \\
 3 &   0.1014 &   0.79 & $-$0.210 & $-$0.399 & 0.917 &   0.711 \\
 4 &   0.1520 &   0.10 & $-$0.257 & $-$0.403 & 0.882 &   0.501 \\
 5 &   0.2027 &$-$0.53 & $-$0.311 & $-$0.550 & 0.879 &   0.071 \\
 6 &   0.2534 &$-$0.97 & $-$0.344 & $-$0.500 & 0.883 &$-$0.093 \\
 7 &   0.3041 &$-$1.30 & $-$0.394 & $-$0.451 & 0.882 &$-$0.378 \\
 8 &   0.3547 &$-$1.67 & $-$0.414 & $-$0.549 & 0.883 &$-$0.517 \\
 9 &   0.4054 &$-$2.28 & $-$0.431 & $-$0.551 & 0.917 &$-$0.523 \\
10 &   0.4561 &$-$2.70 & $-$0.440 & $-$0.550 & 0.918 &$-$0.553 \\
\hline
11 &$-$0.0103 &   1.90 & $-$0.251 & $-$0.204 & 0.878 &   0.773 \\
12 &   0.0404 &   1.15 & $-$0.282 & $-$0.301 & 0.879 &   0.782 \\
13 &   0.0910 &   0.39 & $-$0.329 & $-$0.303 & 0.881 &   0.629 \\
14 &   0.1417 &$-$0.36 & $-$0.378 & $-$0.398 & 0.882 &   0.302 \\
15 &   0.1924 &$-$1.01 & $-$0.435 & $-$0.399 & 0.882 &   0.068 \\
16 &   0.2431 &$-$1.26 & $-$0.490 & $-$0.400 & 0.918 &$-$0.461 \\
17 &   0.2938 &$-$1.37 & $-$0.522 & $-$0.451 & 0.878 &$-$0.725 \\
18 &   0.3444 &$-$1.91 & $-$0.549 & $-$0.450 & 0.883 &$-$0.859 \\
19 &   0.3951 &$-$2.66 & $-$0.578 & $-$0.451 & 0.883 &$-$1.025 \\
20 &   0.4458 &$-$2.83 & $-$0.580 & $-$0.499 & 0.882 &$-$1.000 \\
\enddata
\tablenotetext{a}{HST orbital phase, with zero point set by the first exposure.}
\tablenotetext{b}{Focus obtained from UVIS1 model.}
\tablenotetext{c}{Offset between science image and fringe flat (previous orbit), determined via cross correlation of image rows.}
\tablenotetext{d}{Offset (in x) and scale value for fringe flat (in that orbit) determined by {\bf defringe}.}
\tablenotetext{e}{Offset in dispersion/wavelength (scaled) used in the detrending procedure.}
\end{deluxetable}

\clearpage
\ssectionstar{Appendix B -- Jitter parameters}\label{sec:AppB}
\vspace{-0.3cm}

Recent exoplanet work based on STIS data (e.g., \href{https://ui.adsabs.harvard.edu/abs/2019AJ....158...91S/abstract}{Sing et al. 2019}) has suggested that observed variations in flux among multiple stellar exposures (within a given orbit and/or from orbit to orbit) may be correlated with some measures of the pointing (as obtained from the jitter file associated with each science exposure).  
We therefore examined various parameters in the standard (3-second interval) jitter files for the scanned sub-array exposures obtained for programs 15383 and 16442 -- in order to compare both the intra-orbit and the orbit-to-orbit variations in each parameter with the corresponding observed variations in flux.  
[Given the scan speeds ($\sim$1.1-1.3 pix/sec), comparison of higher-resolution jitter data with the ``striping'' observed within the individual spatially scanned spectral images may likewise provide some insight into those variations (e.g., if they are due to variations in scan speed).]
The parameters examined included the dominant and roll values for V2 and V3; the average SI V2 and V3; the roll, limb, and zenith angles; RA and Dec; latitude and longitude; and the projections of the magnetic field along V1, V2, and V3; see below for definitions and general trends.  
We also considered the CCD housing temperature (as listed in the image header) and the predicted focus values (from the UVIS1 model) for that visit.  
The decreasing focus values through each orbital visibility period suggest that the STIS focus would have improved through each orbit.

\begin{figure}
\centering
\includegraphics[scale=0.7, clip = true, trim = 0 0 0 50]{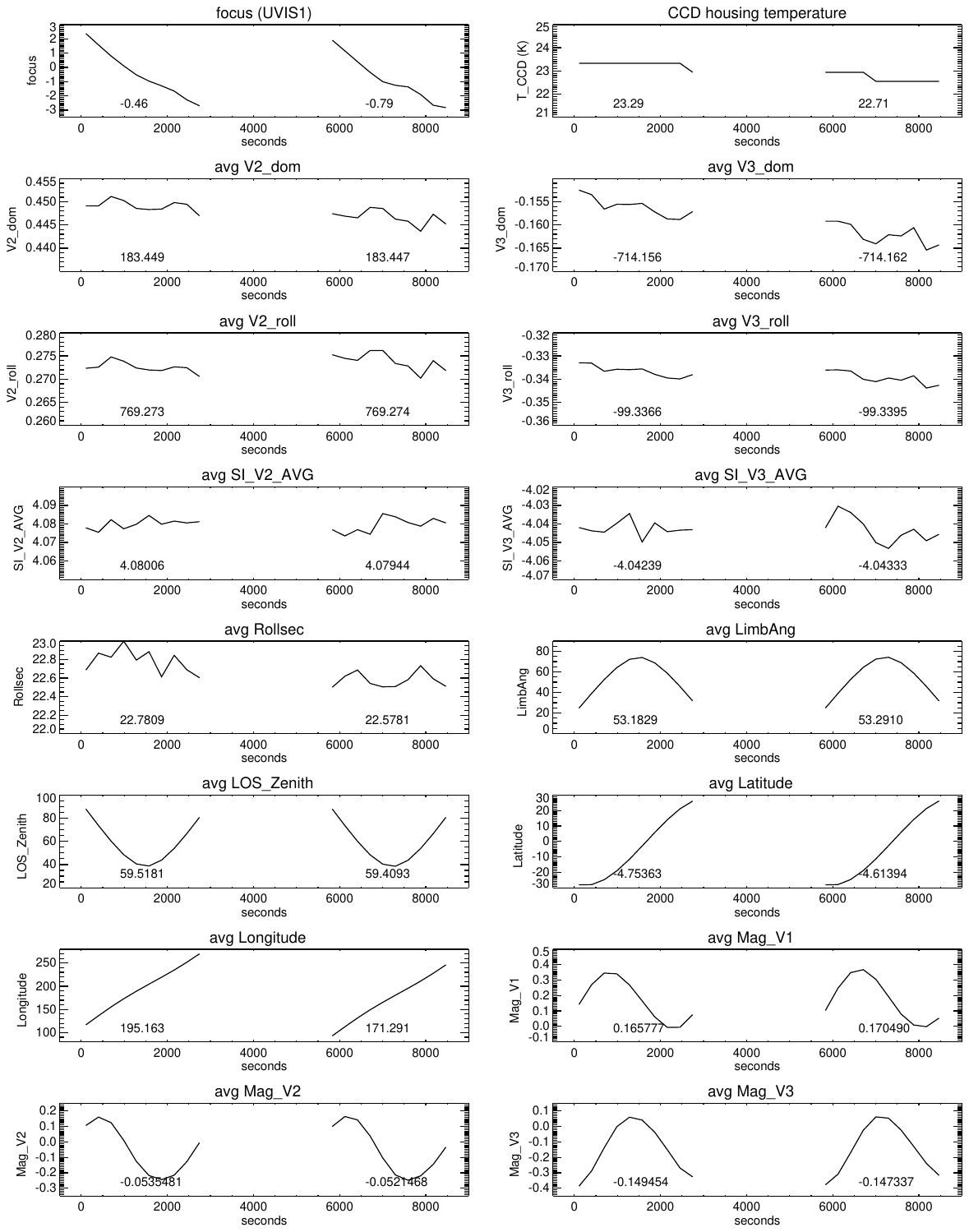}
\caption{{\small Average values of the focus (UVIS1 model), CCD housing temperature, and various jitter parameters for the 20 short scans of 55 Cnc from program 15383 visit 2, as functions of time (with t = 0 at the beginning of the first scan in orbit 2).
The overall average values for the ten scans in each of orbits 2 and 3 are given below the respective curves.
Note that the values for the six V2 and V3 parameters range over about 8 degrees (roughly linearly) during each scanned exposure and that the printed values for the average V2 and V3 dominant and roll parameters include the integers that have been subtracted from the y-axes of the plots.}}
\label{fig:jitavg}
\end{figure}

Table~\ref{tab:jit} lists the average values of each parameter for the 20 sub-array scanned exposures from visit 2 of program 15383; those average values are also shown in Figure~\ref{fig:jitavg}, with overall average values noted for each orbit.  
For the V2 dom, V3 dom, V2 roll, and V3 roll, the average values plotted refer to the decimal part of the values; for the Roll, it is the seconds -- i.e., not the total angles.  
Some general trends in the parameters for the program 15383 scans may be noted:

\begin{itemize}
\item{T = CCD housing temperature -- There is a slight decrease over both orbits, in steps of 0.39.}
\item{focus from UVIS1 model -- The focus decreases over each orbit -- toward better focus for STIS.}
\item{V2 dom, V3 dom = V2, V3 coordinates for dominant FGS (arcsec) -- V2 dom increases linearly from $\sim$179.4 to $\sim$187.8 over each scan; V3 dom decreases linearly from $\sim$$-$710.0 to $\sim$$-$718.4 over each scan.  The average values for both decline slightly over both orbits.}
\item{V2 roll, V3 roll = V2, V3 coordinates for roll FGS (arcsec) -- V2 roll increases linearly from $\sim$765.0 to $\sim$773.8 over each scan; V3 roll decreases linearly from $\sim$$-$95.3 to $\sim$$-$103.5 over each scan; the average values for both decline slightly within each orbit.}
\item{SI V2 avg, SI V3 avg = mean jitter in V2, V3 over 3 seconds (arcsec) -- V2 avg increases linearly from $\sim$0.0 to $\sim$8.4; the average value increases slightly over each orbit.  V3 avg decreases linearly from $\sim$0.0 to $\sim$$-$8.4 over each scan; the average value is roughly constant over each orbit.}
\item{SI V2 rms, SI V3 rms = rms jitter in V2, V3 over 3 seconds (arcsec) -- Both exhibit some variations within each scan and within each orbit, but both are similar between the two orbits.}
\item{Roll = position angle between north and +V3 (degrees) -- The roll is roughly constant (with some scatter) within each scan; it may be slightly lower in orbit 3.}
\item{Limb angle = angle between V1 axis and Earth limb (degrees) -- The limb angle increases from $\sim$25 to $\sim$74, then decreases again to $\sim$32 during each orbit.}
\item{LOS zenith = angle between HST zenith and target (degrees) -- The zenith angle traces part of a roughly sinusoidal variation over each orbit (decreasing from $\sim$88 to $\sim$39, then increasing again), with similar (but perhaps slightly offset) trends in both orbits.}
\item{Latitude, Longitude = latitude and longitude of HST subpoint (degrees) -- The latitude increases from $\sim$$-$28 to $\sim$26, and the longitude increases by $\sim$ 153, over each orbit.}
\item{Mag V1, Mag V2, Mag V3 = magnetic fields along V1, V2, V3 (Gauss) -- The projected fields exhibit (part of) a roughly sinusoidal variation over each orbit, with similar amplitude (0.35-0.40) and phase.}
\end{itemize}

While some of these parameters exhibit trends that bear some similarity to those observed for the raw white-light fluxes (roughly constant within each orbit, but slightly lower in orbit 3 than in orbit 2), dependences on the jitter parameters were not explored in the exploratory fits to the program 15383 data discussed in Sec.~\ref{sec-detrfits}.

\begin{figure}
\centering
\includegraphics[scale=0.7, clip = true, trim = 0 0 0 50]{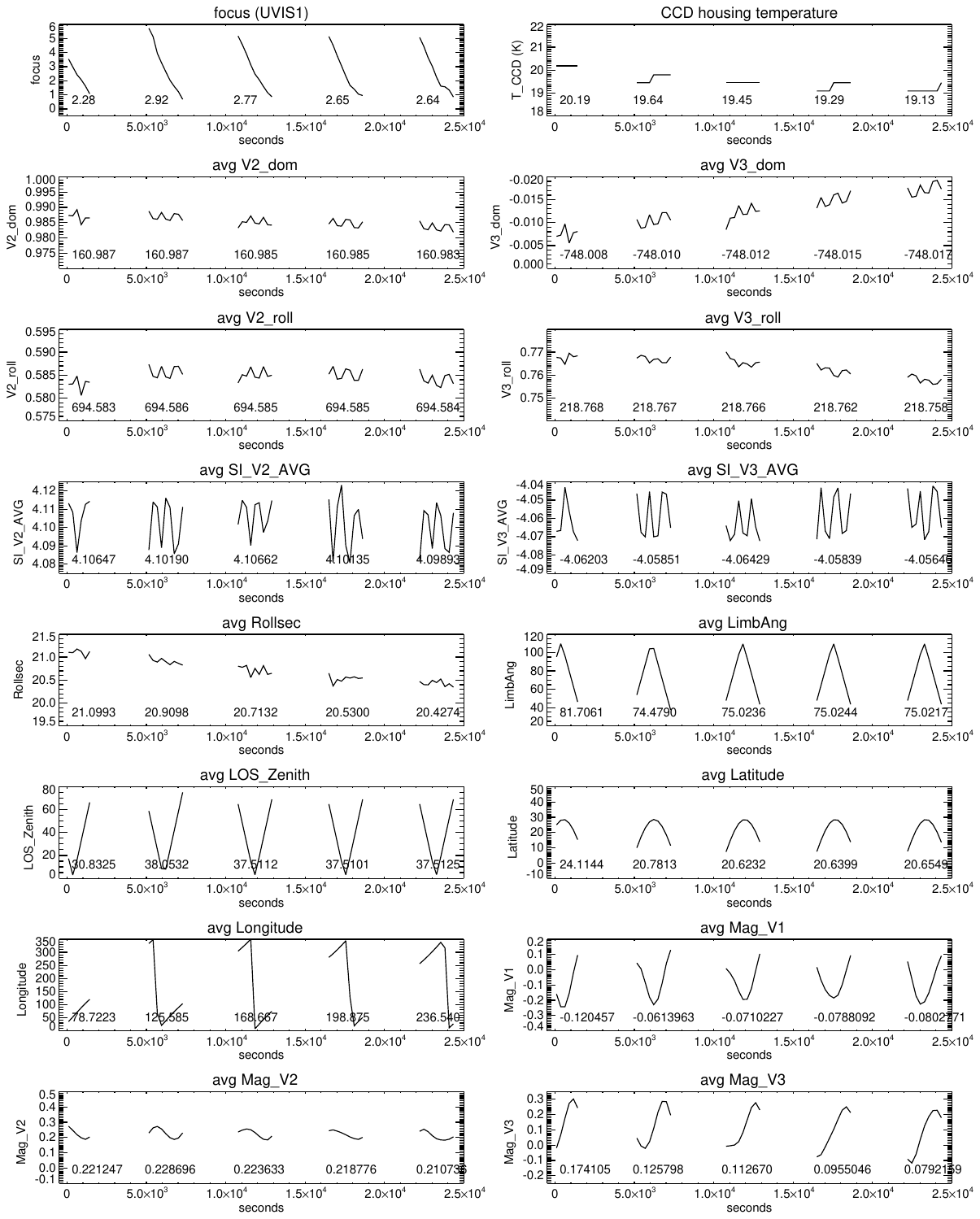}
\caption{{\small Average values of the focus (UVIS1 model), CCD housing temperature, and various jitter parameters for the 42 short scans of 55 Cnc from program 16442, as functions of time (with t = 0 at the beginning of the first scan in orbit 1).
The overall average values for the 6-9 scans in each orbit are given below the respective curves.
Note that the printed values for the average V2 and V3 dominant and roll parameters include the integers that have been subtracted from the y-axes of the plots.}}
\label{fig:jitavg2}
\end{figure}

Figure~\ref{fig:jitavg2} shows the same jitter parameters for all five orbits of program 16442.
Bearing in mind that the observations for the two programs were obtained at different times of the year, the behavior of the various jitter parameters appears to be broadly similar between the two sets of data, in both general trends and scatter (note some differences in the y-axis scales in some panels in the two figures).
After some experimentation, terms in the declination and zenith angle (each to 2nd order) were included in the detrending fits to the fluxes obtained in program 16442 (Sec.~\ref{sec-trspec}).  

\begin{deluxetable}{rrrrrrrrrrrrrrrr}
\rotate
\tablecolumns{16}
\tabletypesize{\scriptsize}
\tablecaption{Average jitter parameters for program 15383 visit 2\label{tab:jit}}
\tablewidth{0pt}

\tablehead{
\multicolumn{1}{c}{Exposure}&
\multicolumn{1}{c}{HST phase\tablenotemark{a}}&
\multicolumn{1}{c}{V2 dom}&
\multicolumn{1}{c}{V3 dom}&
\multicolumn{1}{c}{V2 roll}&
\multicolumn{1}{c}{V3 roll}&
\multicolumn{1}{c}{SI V2 avg}&
\multicolumn{1}{c}{SI V3 avg}&
\multicolumn{1}{c}{Roll}&
\multicolumn{1}{c}{Limb}&
\multicolumn{1}{c}{Zenith}&
\multicolumn{1}{c}{Lat}&
\multicolumn{1}{c}{Long}&
\multicolumn{1}{c}{Mag V1}&
\multicolumn{1}{c}{Mag V2}&
\multicolumn{1}{c}{Mag V3}\\
\multicolumn{1}{c}{(1)}&
\multicolumn{1}{c}{(2)}&
\multicolumn{1}{c}{(3)}&
\multicolumn{1}{c}{(4)}&
\multicolumn{1}{c}{(5)}&
\multicolumn{1}{c}{(6)}&
\multicolumn{1}{c}{(7)}&
\multicolumn{1}{c}{(8)}&
\multicolumn{1}{c}{(9)}&
\multicolumn{1}{c}{(10)}&
\multicolumn{1}{c}{(11)}&
\multicolumn{1}{c}{(12)}&
\multicolumn{1}{c}{(13)}&
\multicolumn{1}{c}{(14)}&
\multicolumn{1}{c}{(15)}&
\multicolumn{1}{c}{(16)}}

\startdata
 1 &   0.0000 &183.449&$-$714.152&769.272&$-$99.333&4.0779&$-$4.0420&22.686&24.783&88.008&$-$28.111&116.683&   0.1422&   0.1064&$-$0.3867\\
 2 &   0.0507 &183.449&$-$714.153&769.273&$-$99.333&4.0755&$-$4.0437&22.869&39.059&73.739&$-$27.994&136.335&   0.2706&   0.1609&$-$0.2881\\
 3 &   0.1014 &183.451&$-$714.157&769.275&$-$99.337&4.0823&$-$4.0445&22.825&52.552&60.229&$-$24.801&155.369&   0.3457&   0.1246&$-$0.1374\\
 4 &   0.1520 &183.450&$-$714.156&769.274&$-$99.336&4.0774&$-$4.0396&22.998&64.221&48.536&$-$19.055&173.033&   0.3402&   0.0100&$-$0.0011\\
 5 &   0.2027 &183.449&$-$714.156&769.272&$-$99.336&4.0798&$-$4.0344&22.794&72.194&40.557&$-$11.535&189.260&   0.2684&$-$0.1264&   0.0602\\
 6 &   0.2534 &183.448&$-$714.155&769.272&$-$99.336&4.0846&$-$4.0498&22.885&73.997&38.738& $-$3.026&204.539&   0.1653&$-$0.2177&   0.0416\\
 7 &   0.3041 &183.448&$-$714.157&769.272&$-$99.338&4.0798&$-$4.0394&22.613&68.799&43.844&    5.737&219.570&   0.0617&$-$0.2447&$-$0.0370\\
 8 &   0.3547 &183.450&$-$714.159&769.273&$-$99.340&4.0815&$-$4.0442&22.846&58.630&53.964&   14.029&235.096&$-$0.0066&$-$0.2153&$-$0.1509\\
 9 &   0.4054 &183.449&$-$714.159&769.272&$-$99.340&4.0806&$-$4.0433&22.690&45.813&66.767&   21.096&251.784&$-$0.0055&$-$0.1282&$-$0.2688\\
10 &   0.4561 &183.447&$-$714.157&769.271&$-$99.338&4.0812&$-$4.0431&22.603&31.782&80.799&   26.123&269.962&   0.0757&$-$0.0050&$-$0.3265\\
\hline
11 &$-$0.0103 &183.447&$-$714.159&769.275&$-$99.336&4.0769&$-$4.0421&22.500&24.833&87.958&$-$28.148& 92.872&   0.1030&   0.0999&$-$0.3780\\
12 &   0.0404 &183.447&$-$714.159&769.274&$-$99.336&4.0735&$-$4.0304&22.620&39.144&73.652&$-$27.950&112.517&   0.2485&   0.1646&$-$0.3089\\
13 &   0.0910 &183.447&$-$714.160&769.274&$-$99.337&4.0769&$-$4.0338&22.687&52.683&60.097&$-$24.683&131.529&   0.3483&   0.1425&$-$0.1690\\
14 &   0.1417 &183.449&$-$714.163&769.276&$-$99.340&4.0745&$-$4.0400&22.542&64.392&48.364&$-$18.884&149.148&   0.3676&   0.0378&$-$0.0241\\
15 &   0.1924 &183.449&$-$714.164&769.276&$-$99.341&4.0856&$-$4.0501&22.505&72.400&40.349&$-$11.324&165.353&   0.3047&$-$0.1039&   0.0619\\
16 &   0.2431 &183.446&$-$714.162&769.273&$-$99.340&4.0839&$-$4.0532&22.509&74.193&38.541& $-$2.802&180.611&   0.1929&$-$0.2141&   0.0537\\
17 &   0.2938 &183.446&$-$714.162&769.273&$-$99.341&4.0808&$-$4.0461&22.582&68.944&43.698&    5.955&195.642&   0.0795&$-$0.2484&$-$0.0238\\
18 &   0.3444 &183.444&$-$714.161&769.270&$-$99.339&4.0788&$-$4.0429&22.734&58.714&53.880&   14.226&211.189&   0.0087&$-$0.2202&$-$0.1297\\
19 &   0.3951 &183.447&$-$714.165&769.274&$-$99.344&4.0830&$-$4.0491&22.592&45.844&66.736&   21.252&227.913&$-$0.0020&$-$0.1459&$-$0.2389\\
20 &   0.4458 &183.445&$-$714.164&769.272&$-$99.343&4.0806&$-$4.0456&22.511&31.763&80.818&   26.218&246.133&   0.0536&$-$0.0340&$-$0.3165\\
\enddata
\end{deluxetable}


\end{document}